\documentclass[12pt,epsf]{article}
  \usepackage{epsfig}
  \newlength{\abstractwidth}
  \renewcommand{\thefootnote}{\fnsymbol{footnote}}
  \renewcommand{\thanks}[1]{\footnote{#1}} 
  \newcommand{\starttext}{
  \setcounter{footnote}{0}
  \renewcommand{\thefootnote}{\arabic{footnote}}}
  \renewcommand{\theequation}{\thesection.\arabic{equation}}
  \newcommand{\be}{\begin{equation}}
  \newcommand{\bea}{\begin{eqnarray}}
  \newcommand{\eea}{\end{eqnarray}}
  \newcommand{\beq}{\begin{equation}}
  \newcommand{\ee}{\end{equation}}
  \newcommand{\eeq}{\end{equation}}
  
  \newcommand{\<}{\langle\,}

  \renewcommand{\>}{\rangle}

  \def\ba{\begin{eqnarray}}
  \def\ea{\end{eqnarray}}

\def\wdw{Wheeler-DeWitt}

  \def\12{{1 \over 2}}
  \def\eq{&=&}

  \def\h3{h^{3\over 2}}

  \def\cc{cosmological constant }
  
  \def\simleq{\; \raise0.3ex\hbox{$<$\kern-0.75em
      \raise-1.1ex\hbox{$\sim$}}\; }
   \def\simgeq{\; \raise0.3ex\hbox{$>$\kern-0.75em
      \raise-1.1ex\hbox{$\sim$}}\; }

\def\sig{$\Sigma$}
\def\ads{anti de Sitter space}
\def\cft{conformal field theory}
\def\cdl{Coleman De Luccia}

\def\ba{\bf{a}}
  
\begin{document}
  \renewcommand{\theequation}{\thesection.\arabic{equation}}
  \begin{titlepage}
  \bigskip

  \bigskip\bigskip\bigskip\bigskip

  \centerline{\Large \bf {The Census Taker's Hat}}

  \bigskip \bigskip

  \bigskip\bigskip
  \bigskip\bigskip

\begin{center}
  {{\large Leonard Susskind
  \footnote{Lecture given at ``String Phenomenology and Cosmology Workshop,"
KIAS and Yukawa Institute, September 24-28, 2007 }
  }}
  \bigskip
  \\ School of Physics, Korea Institute for Advanced Study (KIAS), Seoul 130-722, Korea,

  \bigskip
  and

\bigskip
   Department of Physics,
  Stanford University\\ Stanford, CA 94305-4060, USA \\ \vspace{2cm}
  \end{center}

  \bigskip\bigskip
  \begin{abstract}




If the observable universe  really is a hologram, then of
what sort? Is it rich enough to keep track of an eternally
inflating multiverse? What physical and mathematical principles
underlie it? Is the hologram a lower dimensional quantum field
theory, and if so, how many dimensions are explicit, and how many
``emerge?" Does the Holographic description provide clues for
defining a probability measure on the Landscape?

The purpose of this lecture is first, to briefly review   a
 proposal  for a holographic cosmology by Freivogel, Sekino, Susskind, and Yeh
(FSSY), and then to develop a physical interpretation in terms of
a ``Cosmic Census Taker:" an idea introduced  in reference [1].
The  mathematical structure--a hybrid of  the Wheeler DeWitt
formalism and holography--is a boundary ``Liouville" field theory,
whose UV/IR duality is closely related to the  time evolution of
the Census Taker's observations. That time evolution is
represented by the renormalization-group flow of the Liouville
theory.

Although quite general, the Census Taker idea was originally
introduced in \cite{shenker}, for the purpose of counting  bubbles
that collide with the Census Taker's bubble. The "Persistence of
Memory" phenomenon discovered by Garriga, Guth, and Vilenkin,  has
a natural RG interpretation,  as does slow roll inflation.  The RG
flow and the related C-theorem are closely connected with
generalized entropy bounds.

  \medskip
  \noindent
  \end{abstract}

  \end{titlepage}
  \starttext \baselineskip=17.63pt \setcounter{footnote}{0}


\setcounter{equation}{0}
  \section{ Introduction }

Of all the  ``String Inspired" cosmological scenarios,   only one
seems to me to have
 an element of inevitability to it. The facts and
principles that drive it are as follows:
\bigskip

\begin{itemize}

\item Observational evidence supports the existence of a period of
slow-roll inflation during which the universe exponentially
expanded by a factor no less than $e^{50}$. The universe grew to a
size which is at least $1,000$ times larger (in volume) than the
portion which is observable.

\item A small residual vacuum energy of order $10^{-123}M_p^4$
remained at the end of inflation and now dominates the energy
density of the universe. If this situation persists, then not only
is the universe at least $1,000$ larger than what can be seen; it
is $1,000$ larger than what can  \textbf{\emph{ever}}  be seen
\cite{matt}.

 \item String Theory apparently gives rise to an immense Landscape of de Sitter vacua
\cite{bousso-polch}\cite{landscape}\cite{kklt}\cite{douglas} with
a very dense ``discretuum" of vacuum energies. None of these vacua
are absolutely stable: each can decay to vacua with smaller \cc.
\item Black Hole (or Observer) Complementarity,
\cite{complementarity} \cite{verlinde} \cite{tom willy}, and the
Holographic Principle, \cite{hologram} have been confirmed by
string theory, at least in a certain wide class of backgrounds
\cite{juan}\cite{witten}\cite{susskind and witten}. The
implication is twofold. On the one hand, observer complementarity
requires the identification of a causal patch; conventional
quantum mechanics only makes sense within such a patch. The
Holographic Principle requires that a region of space be described
by boundary degrees of freedom whose number does not exceed the
area, measured in Planck Units.

\item Inflation, if it lasts long enough, has a tendency
\cite{eternal} \ to populate the Landscape with a great diversity
of nucleated ``pocket universes."

\end{itemize}
\bigskip

The first two items imply that all of observable cosmology
consisted of a roll from one value of the vacuum energy (probably
no bigger than $~10^{-14}M_p^4$), to its final current value. How
and why the universe began with such an unnatural energy density
is not explained by any standard theory, but the Landscape
suggests  the following guess: At some point in the remote past
the universe occupied a point on the Landscape with a much higher
vacuum energy, perhaps of order one in Planck units. Rolling,
unimpeded, to a vacuum energy of $10^{-14} $ without getting stuck
in a local minimum is unlikely. (Think of rolling   a bowling ball
from the top of Mount Everest to sea level.)  It is far more
likely that the  universe  would get stuck in many minima, and
have to tunnel \cite{Coleman} multiple times, before arriving at
the very small vacuum energy required by conventional slow-roll
inflation. We will not dwell on Anthropic issues in this paper,
but I would point out that a long period of conventional inflation
appears to be required for structure formation \cite{maria}. The
argument is similar to the well-known Weinberg argument concerning
the cosmological constant.

\bigskip

These considerations strongly suggest that the period of
conventional slow-roll inflation was preceded by a tunneling event
from a previous neighboring vacuum. In other words, the observed
universe evolved by a sudden bubble nucleation from an
\textbf{\emph{``Ancestor"}} vacuum, once removed on the Landscape.
It seems obvious  that one of the next big questions for cosmology
will be to find the theoretical and observational tools to confirm
or refute the past existence of an Ancestor, and to find out as
much as we can about it. If we are lucky and the amount of
slow-roll inflation that followed bubble nucleation is as small as
observational evidence allows, then we have a chance of seeing
features of the Ancestor imprinted on the sky \cite{maria}. The
two smoking guns would be:

\begin{itemize}

\item Negative spatial curvature: bubble nucleation leads to a
negatively curved, infinite,  FRW universe.

\item Tensor modes in the CMB, but only in the lowest harmonics.
Although the vacuum energy subsequent to tunneling (during
conventional slow-roll inflation) was almost certainly too small
to create observable tensor modes, the \cc \ in the Ancestor was
probably much larger. During the Ancestor epoch, large tensor
fluctuations would be created by rapid inflation. A tail
(diminishing rapidly with $l$ ) of those fluctuations could be
visible if the number of slow-roll e-foldings is minimal.

\end{itemize}

\bigskip

If the observational evidence for an Ancestor is weak, so is the
current theoretical framework. To many of us, eternal inflation,
bubble nucleation, and a multiverse,  seem all but inevitable, but
it is also true that they have inspired what
Bjorken\footnote{James Bjorken, private communication.} has called
``the most extravagant extrapolation in the history of physics."
Eternal inflation leads to an uncontrolled infinity of ``pocket
universes" which we have no good idea how to regulate--the
inevitable  has led  to the preposterous. In my opinion, this
situation reflects  serious confusion, and perhaps even a crisis.

\bigskip

Eternal inflation is not the only extravagance that we have had to
tame in recent decades. I have in mind the fact that a naive but
very compelling interpretation of black holes seemed, at one time,
to imply that a black hole can absorb an infinite amount of
information behind its horizon \cite{hawking}. By feeding a black
hole with coherent energy at the same rate that it evaporates, it
would seem that an infinity of bits could be lost to the
observable world.

\bigskip

I believe these two crises may be related. In both cases the
infinities result from ``cutting across horizons" and attempting
to describe global space-like surfaces with
\textbf{\emph{independent}}   degrees of freedom at each location.
The cure is to focus attention on a single causal region, and to
describe it by a Holographic set of degrees of freedom
\cite{complementarity} \cite{inside raphael}.

\bigskip

In FSSY \cite{yeh} the authors described one such holographic
framework--call it holography in a hat--based on mathematical
ideas that have become familiar from String Theory. At the same
time, Shenker and collaborators \cite{shenker} have developed an
intuitive ``gedanken observational" approach based on a fictitious
observer called the ``Census Taker."  My purpose in this lecture
is to explain the  close connection between these ideas.

\setcounter{equation}{0}
  \section{The Census Bureau}

  Let us begin with a precise definition of a causal patch. Start with a cosmological
space-time and assume that a future causal boundary exists. For
example, in flat Minkowski space the future causal boundary
consists of ${\cal{I}}^+$ (future light like infinity) and a
single point, time-like-infinity.
  For a (non-eternal) Schwarzschild black hole, the future causal boundary has an
additional component: the  singularity.
\bigskip

  A causal patch is defined in terms of a point $\bf\underline{a}$ on the future causal
boundary. I'll call that point the ``Census Bureau\footnote{This
term originated during a discussion between myself and Steve
Shenker in a Palo Alto Cafe. Neither of us will admit to having
coined it first, but it wasn't me.}."  The causal patch is, by
definition, the causal past of the Census Bureau,  bounded by its
past light cone. For Minkowski space, one usually  picks the
Census Bureau to be time-like-infinity. In that case the causal
patch is all of Minkowski space as seen in figure 1.

 \bigskip
 \begin{figure}
\begin{center}
\includegraphics[width=12cm]{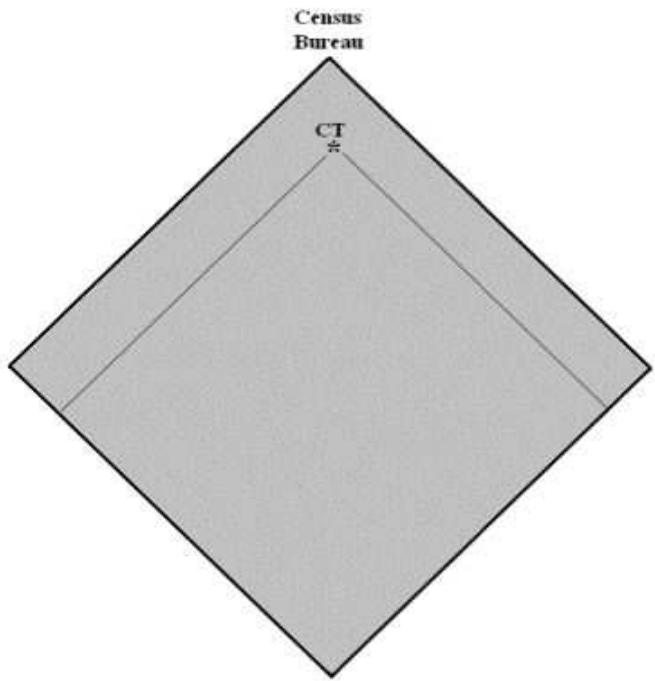}
\caption{Conformal diagram for ordinary flat Minkowski space. The
 causal patch associated with the ``Census Bureau" is the entire space-time. A Census
 Taker and his past light-cone are also shown.} \label{1}
\end{center}
\end{figure}
 \bigskip

  \bigskip

\bigskip

  In the case of the Schwarzschild geometry, $\bf\underline{a}$ can again be chosen to be
time-like-infinity, in which case the causal patch is everything
outside the horizon of the black hole. There is no clear reason
why one can't choose $\bf\underline{a}$ to be on the singularity,
but it  would lead to obvious difficulties.

\bigskip

The term ``Census Taker" was introduced \cite{shenker}) to denote
an observer, at a point inside a causal patch, who looks back into
the past and collects data. He can count galaxies, other
observers, hydrogen atoms, colliding bubble-universes,
civilizations, or anything else within his own causal past. As
time elapses the Census Taker sees more and more of the causal
patch. Eventually all  Census takers within the  causal patch
arrive at the Census Bureau where they can compare data.

  \bigskip
 \begin{figure}
\begin{center}
\includegraphics[width=12cm]{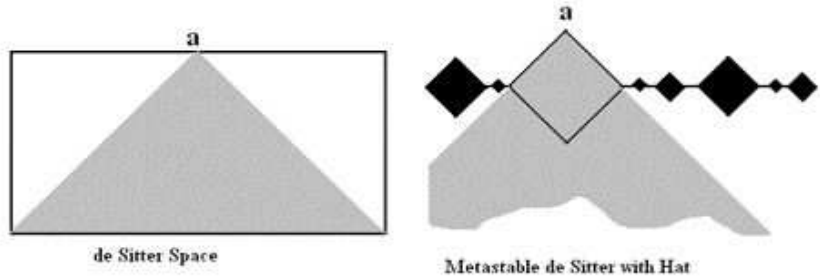}
\caption{Conformal diagrams for eternal and metastable de Sitter
space. The grey areas are causal patches associated with the
points $\bf{a}$. In the metastable case the causal patch is
associated with the tip of a hat.} \label{2}
\end{center}
\end{figure}
 \bigskip

De Sitter space has the well known causal structure as shown in
figure 2. In this case all  points at  future infinity are
equivalent: the Census Bureau can be located at any of them.
However, String Theory and other considerations suggest that  de
Sitter minima are never stable. After a series of tunneling events
they eventually end in
 terminal vacua with exactly zero or negative cosmological
constant. The entire distant future of de Sitter space is replaced
by a fractal of terminal bubbles.

\bigskip

Decay to negative cosmological constant always leads to a singular
crunch. Barring governmental stupidity, this seems an unlikely
place for a Census Bureau. The disadvantages (or advantages) of
locating a government agency  at a crunch are the same as at a
black hole singularity.

\bigskip

Terminal vacua with zero \cc \ seem more promising;  the bubble
then evolves to an open, negatively curved, FRW geometry, bounded
by a ``hat" \cite{yeh}. The Census Bureau is at the tip of the
hat.

\bigskip

  In the case of the black hole, the degrees of freedom beyond the horizon, i.e., outside
the causal patch, are redundant descriptions of degrees of freedom
within the patch: they should not be double-counted. We assume
that the same is true of the causal patch of a hat. In both cases
the conventional rules of quantum mechanics are expected to apply
\textbf{\emph{only}} within the causal patch.  Furthermore the
rules should respect the Holographic Principle.

  \bigskip

  The reader may wonder about the relationship between hatted terminal geometries, and
observational cosmology with a non-zero cosmological constant.
There are two answers: the first is that for many purposes, the
current \cc \ is so small that  it can be set to zero. Later we
will argue that the \cft \ description of the approximate hat
which results from non-zero cosmological constant is an
ultraviolet cut-off version of the type of field theory that
describes a hat.

  \bigskip

  The second answer was emphasized by Shenker, et al. \cite{shenker} \ who argued that
because our present de Sitter vacuum will eventually decay, a
Census Taker can look back into our current vacuum from a point at
or near the tip of a hat, and gather information. In principle the
Census Taker can peek back, not only into the Ancestor vacuum (our
vacuum in this case), but also into bubble collisions with other
vacua of the Landscape. Much of this paper is about the gathering
of information as the Census Taker's time progresses, and how it
is encoded in the renormalization-group (RG) flow of a holographic
field theory.

\setcounter{equation}{0}
  \section{Open FRW and Euclidean ADS }

The classical space-time in the interior of a \cdl \ bubble,  has
the form of an open infinite FRW universe, Let  ${\cal{H}}_3$
represent a hyperbolic geometry with constant negative curvature.
\be d {\cal{H}}_3^2 = dR^2 + \sinh^2{R} \ \ d\Omega_2^2.
\label{hyperbolic} \ee The metric of open FRW is \be ds^2 = -dt^2
+ a(t)^2 d{\cal{H}}_3^2, \label{FRW metric in terms of T} \ee or
in terms of  conformal time $T$ (defined by $dT = dt/ a(t) $) \be
ds^2 = a(T)^2 ( -dT^2 + d{\cal{H}}_3^2) \label{FRW metric in terms
of T} \ee Note that in (\ref{hyperbolic}) the radial coordinate
$R$ is a hyperbolic angle and that the symmetry of the spatial
sections is the non-compact group $O(3,1)$.  This $O(3,1)$
symmetry  plays  a central  role in what follows.

\bigskip

If the vacuum energy in the bubble is zero, i.e., no cosmological
constant, then the future boundary of the FRW region is a hat. The
scale factor $a(t) $ then has the  early and late-time behaviors
\bea a(T) \sim t \sim   De^T. \label{asymptotic a} \eea.

For early time when $T \to -\infty$ the constant $D$ is
conveniently chosen to  be the Hubble scale of the Ancestor,
$H^{-1}$. \be a(T) = H^{-1} e^T \ \ \  \ \ \rm (T \to - \infty)
\ee At late time it is always larger.  In  the  simplest thin-wall
case $D$ is given by  the Ancestor Hubble-length at all times.

\bigskip

\begin{figure}
\begin{center}
\includegraphics[width=12cm]{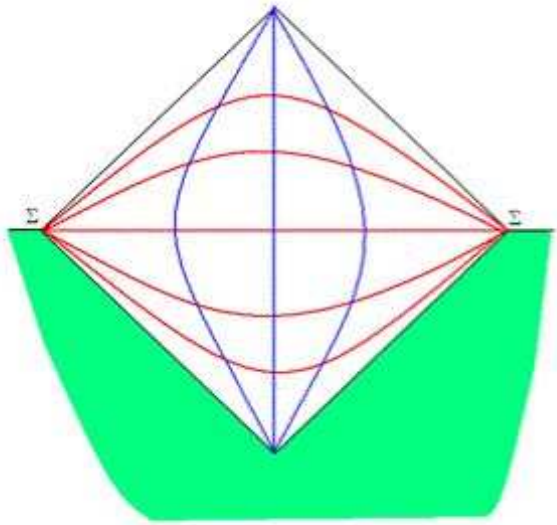}
\caption{A Conformal diagram for the FRW universe created by
bubble nucleation from an ``Ancestor" metastable vacuum. The
Ancestor vacuum is shown in green. The red and blue curves are
surfaces of constant $T$ and $R$. The two-sphere at spatial
infinity is indicated by \sig.} \label{3}
\end{center}
\end{figure}
 \bigskip

  In figure 3,
a conformal diagram of FRW is illustrated,  with surfaces of
constant $T$ and
 $R$  shown in red and blue. The green region
represents the de Sitter Ancestor vacuum. Figure 4 shows the the
Census Taker, as he approaches the tip of the hat, looking back
along his past light cone.

 \begin{figure}
\begin{center}
\includegraphics[width=12cm]{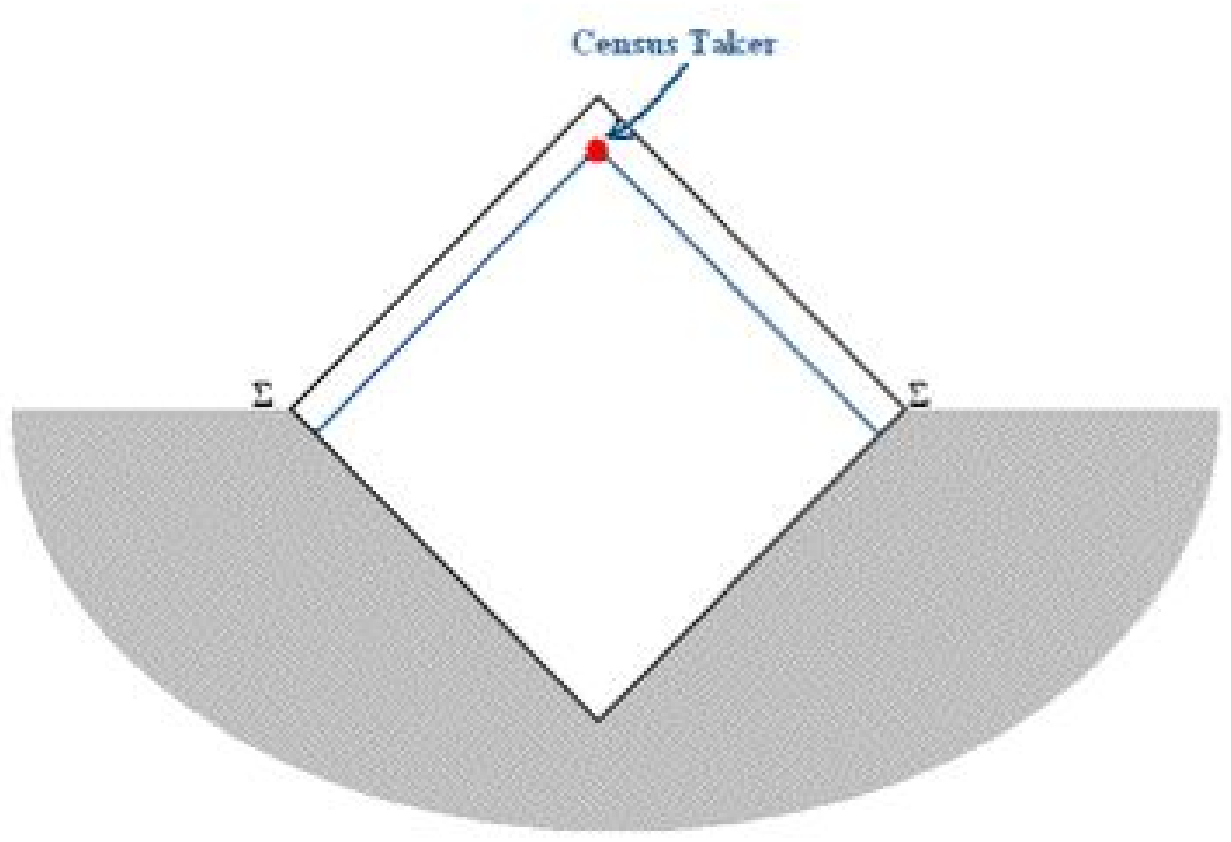}
\caption{The Census Taker is indicated by the red dot. The blue
lines represent his past light-cone. } \label{4}
\end{center}
\end{figure}
 \bigskip

\bigskip

Part of the inspiration for FSSY was the geometry of the spatial
slices of constant $T$. Each slice, taken by itself, is a three
dimensional, negatively curved, hyperbolic plane. It is very
familiar to relativists and string theorists, being identical to
3-D Euclidean \textbf{\emph{anti de Sitter space}}. The best way
that I know of for becoming familiar with the hyperbolic plane is
to study Escher's  drawing ``Limit Circle IV." It is both a
drawing of Euclidean ADS and also a fixed-time slice of open FRW.
 In figure 5,
the green circle is the intersection of Census Takers past light
cone with the time-slice. As the Census Taker advances in time,
the green circle moves out, ever closer to the boundary.

\begin{figure}
\begin{center}
\includegraphics[width=12cm]{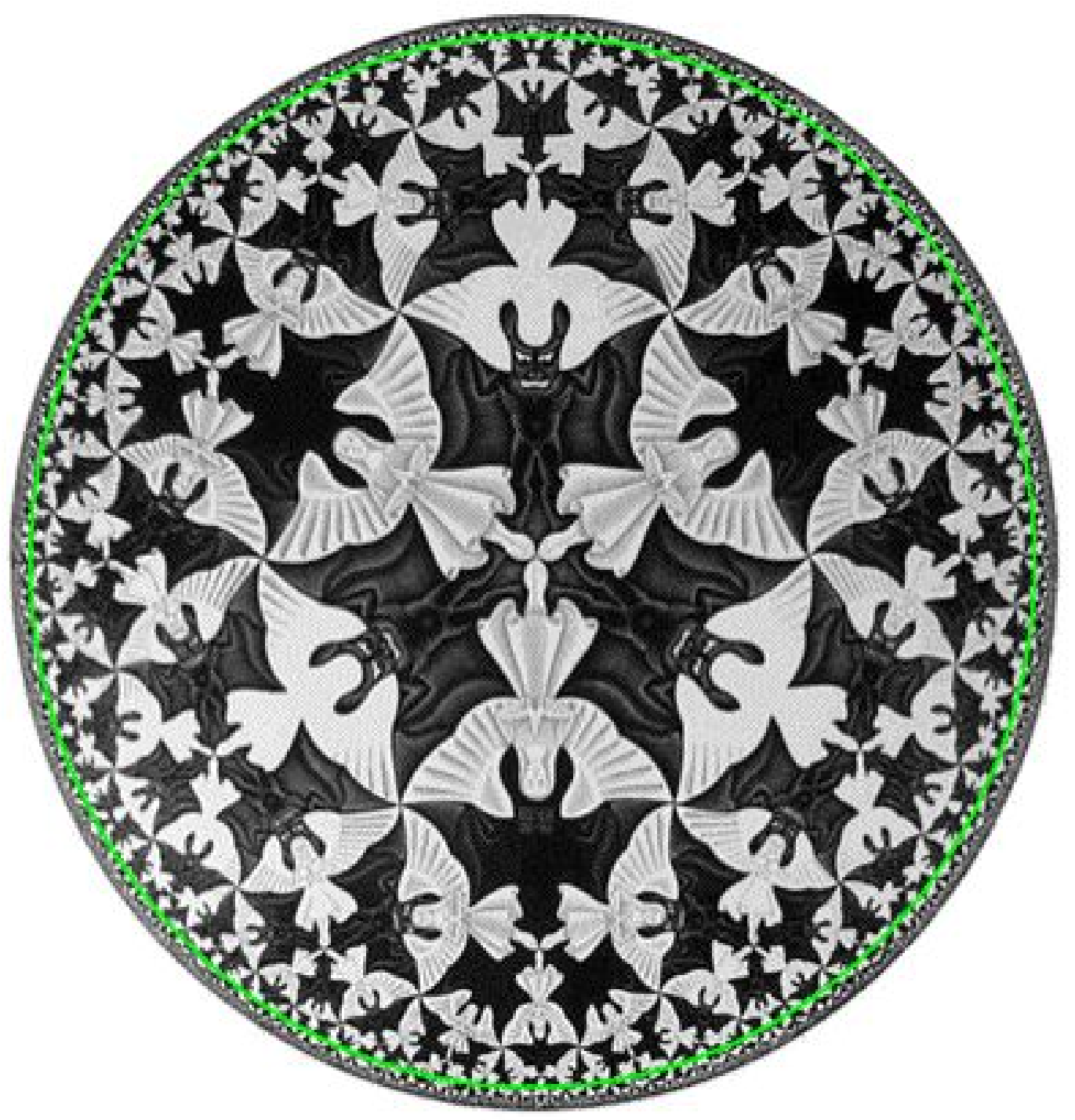}
\caption{Escher's drawing of the Hyperbolic Plane, which
represents Euclidean \ads \ or a spatial slice of open FRW. The
green circle shows the intersection of the Census Taker's past
light-cone, which moves toward the boundary with
Census-Taker-time.} \label{5}
\end{center}
\end{figure}
 \bigskip

A fact (to be explained later) which will play a leading role in
what follows, concerns the Census Taker's angular resolution,
i.e., his ability to discern small angular variation. If the time
at which the CT looks back is called $T_{CT} $, then the smallest
angle he can resolve  is of order $\exp{(-T_{CT})}$. It is as if
the CT were looking deeper and deeper into the ultraviolet
structure of a quantum field theory on \sig.

\bigskip

The boundary of \ads \ plays a key role in the ADS/CFT
correspondence, where it represents the extreme ultraviolet
degrees of freedom of the boundary theory. The corresponding
boundary in the FRW geometry is labeled \sig \ and consists of the
intersection of the hat ${\cal{I}}^+$, with the space-like future
boundary of de Sitter space. From within the interior of the
bubble, \sig  \ represents \textbf{\emph{ space-like infinity.  }}
It is the obvious surface for a holographic description. As one
might expect, the $O(3,1)$ symmetry which acts on the time-slices,
also has the action of two dimensional conformal transformations
on \sig.  Whatever the Census Taker sees, it is very natural for
him to classify his observations under the conformal group. Thus,
the apparatus of (Euclidean) conformal field theory, such as
operator dimensions, and correlation functions, should play a
leading role in organizing his data.

\bigskip

In complicated situations, such as multiple bubble collisions,
\sig \ requires a precise definition. The asymptotic light-cone
${\cal{I}}^+$ (which is, of course, the limit of the Census Takers
past light cone), can be thought of as being formed from a
collection of light-like generators.  Each generator, at one end,
runs into the tip of the hat, while the other end eventually
enters the bulk space-time. The set of points where the generators
enter the bulk define \sig.

\setcounter{equation}{0}
\section{The Holographic Wheeler DeWitt Equation}

Supposedly, String Theory is a quantum theory of gravity, and
indeed it has proved to be a remarkably powerful one, but only in
certain special  backgrounds. As effective as it is in describing
scattering amplitudes in flat (supersymmetric) space-time, and
black holes in \ads,  it is an inflexible tool which at present is
close to useless for formulating a mathematical  framework for
cosmology. What is it that is so special about flat and ADS space
that allows a rigorous formulation of quantum gravity, and why are
cosmological backgrounds so difficult?

\bigskip

The problem is  frequently blamed on
\textbf{\emph{time-dependence}}. But time-dependent deformations
of \ads \ or Matrix Theory are easy to describe. Something else is
the culprit. There is one important difference between the usual
String Theory backgrounds and  more interesting  cosmological
backgrounds. Asymptotically-flat and anti de Sitter backgrounds
have a property that I will call \textbf{\emph{asymptotic
coldness}}.  Asymptotic coldness  means  that the boundary
conditions require the energy density to go to zero at the
asymptotic boundary of space\footnote{Note that asymptotic
coldness refers only to conditions at spatial infinity. A
violation of asymptotic coldness does not imply that the
temperature remains finite as the time goes to infinity, although
even this is a problem in geometries that contain de Sitter
boundary conditions. Hats are somewhat better in that they become
cold at late time.}. Similarly, the fluctuations in geometry
tend to zero.  This condition is embodied in the statement that
all physical disturbances are composed  of normalizable modes.
Asymptotic coldness is obviously important to defining an S-matrix
in flat space-time, and plays an equally important role in
defining the observables of \ads.

\bigskip

But in cosmology, asymptotic coldness is  never the case. Closed
universes have no asymptotic boundary,  and homogeneous infinite
universes have matter, energy, and geometric variation out to
spatial infinity; under the circumstances an S-matrix cannot be
formulated. String theory at present is ill equipped to deal with
\textbf{\emph{asymptotically warm}} geometries. To put it another
way, there is a conflict between a homogeneous cosmology, and the
Holographic Principle which requires an isolated, cold, boundary.

\bigskip

The traditional approach to quantum cosmology--the \wdw \
equation--is the opposite of string theory; it is
 very flexible from the point of
view of background dependence--it doesn't require any definite
boundary condition, it can be formulated for a closed universe, a
flat or open FRW universe,  de Sitter space, or for that matter,
flat and \ads \ spacetime--but it is not
 a consistent quantum theory of gravity. It is based on an
obsolete approach--local quantum field theory--that fails to
address the problems that String Theory and the  Holographic
Principle were designed to solve: the huge over-counting of
degrees of freedom implicit in a local field theory.

\bigskip

FSSY suggested a way out of dilemma: synthesize the \wdw \
philosophy with the Holographic Principle to construct a
Holographic \wdw \ theory. We will begin with a review of the
basics of conventional WDW; For a more complete treatment,
especially of infinite cosmologies, see \cite{banks et al}.

\bigskip

The ten equations of General Relativity take the form \be {\delta
\over \delta g_{\mu \nu}  } I =0 \label{GR equations} \ee where
$I$ is the Einstein action for gravity coupled to matter. The
canonical formulation of General Relativity makes use of a
time-space split \cite{ADM}. The six space-space components are
more or less conventional equations of motion, but the four
equations involving the time index have the form of constraints.
These four equations are written,
 \be
 H^{\mu}(x) =0.
 \label{adm}
 \ee
 They involve the space-space components of the metric $g_{nm}$, the matter fields
$\Phi$, and their conjugate momenta. The time component $H^0(x)$,
is a local Hamiltonian
 which ``pushes time forward" at the spatial point $x$. More generally, if  integrated
with a test function,
 \be
 \int d^3x \ f(x) \ H^0(x)
 \label{local push generator}
 \ee
it generates infinitesimal transformations of the form \be t \to t
\ + \ f(x).
 \label{local push}
\ee

\bigskip

Under certain conditions  $H^{0}$ can be integrated over space in
order to give a global Hamiltonian description. Since $H^{0}$
involves second space derivatives of $g_{nm}$, it is necessary to
integrate by parts in order  to bring the Hamiltonian to the
conventional form containing only first derivatives. In that case
the ADM equations can be written as \be \int d^3x \ H = E.
\label{H equals E} \ee The Hamiltonian density $H$ has a
conventional structure, quadratic in canonical momenta, and the
energy $E$ is given by a Gaussian surface integral over  spatial
infinity. The conditions which allow us to go from (\ref{adm})  to
(\ref{H equals E}) are satisfied in asymptotically cold
flat-space-time, as well as in \ads; in both cases global
Hamiltonian formulations exist. Indeed, in \ads \ the Hamiltonian
of the Holographic boundary description is identified with the ADM
Energy, but, as we noted, cosmology, at least in its usual forms,
is never asymptotically cold. The only recourse for a canonical
description, is the local form of the equations (\ref{adm}).

\bigskip

When we pass from classical gravity to its quantum counterpart,
the usual generalization of the canonical equations (\ref{adm})
become the Wheeler DeWitt equations, \be H^{\mu} |\Psi \> = 0
\label{ H^mu Psi =0} \ee where the state vector $|\Psi \>$ is
represented by a wave functional that depends only on  the space
components of the metric $g_{mn}$, and the matter fields $\Phi$.

\bigskip

The first three equations \be H^{m} |\Psi \> = 0 \ \ \ (m = 1, \
2, \ 3.) \label{momentum constraits} \ee have the interpretation
that the wave function is invariant under spatial diffeomorphisms,
 \be
 x^n \to x^n + f^n(x^m)
 \label{space diffeos}
 \ee
 In other words $\Psi(g_{mn}, \ \Phi)$ is a function of spatial invariants. These
equations are usually deemed to be the easy \wdw \ equations.

 \bigskip

The difficult equation is the time component \be H^{0} |\Psi \> =
0. \label{hamiltonian constraint} \ee It represents invariance
under local, spatially varying, time translations. Not only is
equation (\ref{hamiltonian constraint}) difficult to solve; it is
difficult to  even formulate: the expression for $H^0$ is riddled
with factor ordering ambiguities. Nevertheless, as long as the
equations are not pushed into extreme quantum environments, they
can be useful.

\bigskip

\subsection{Wheeler DeWitt and the Emergence of Time}

Asymptotically cold backgrounds come equipped with a global
concept of time. But in the more interesting asymptotically warm
case, time is an approximate, derived, concept \cite{banks et
al,banks}, which emerges from the solutions to the \wdw \
equation. The perturbative method for solving (\ref{hamiltonian
constraint}) that was outlined in \cite{banks et al}, can be
adapted to the case of negative spatial curvature. We begin by
decomposing the spatial metric  into a constant curvature
background, and fluctuations. Since we will focus on open FRW
cosmology, the spatial curvature is negative, the space metric
having the form,
 \be ds^2 = a^2 \left(dR^2 + \sinh^2 R \ \ (d\theta^2 +
\sin^2 \theta d\phi^2) \right) + a^2 h_{mn}dx^m dx^n
\label{complete spatial frw metric} \ee In (\ref{complete spatial
frw metric})\ $a$ is the usual FRW scale factor and the $x's$  are
$(R, \ \ \theta, \ \phi)$.

\bigskip

The first approximation, in which all fluctuations are ignored,
 is usually called the mini-superspace approximation, but
it really should be seen as a first step in a semiclassical
expansion. In lowest order, the Wheeler DeWitt wave function
depends only on the scale factor $a$. To carry out the leading
approximation in open FRW it is necessary to introduce an infrared
regulator which can be done by bounding the value of $R$, \be R <
R_0 \ \ (R_0 >> 1). \label{R=IR cutoff} \ee Lets also define the
total dimensionless coordinate-volume within the cutoff region, to
be $V_0$. \be V_0 = 4 \pi \int dR \sinh^2R \approx \12\pi
e^{2R_0}. \label{IR cutoff volume} \ee

\bigskip

The first (mini-superspace) approximation is described by the
action, \be L = {-a V_0\dot{a}^2 - V_0a   \over 2} \label{miniL}
\ee Defining  $P$ to be the momentum conjugate to the scale factor
$a$, \be P = -a V_0 \dot{a} \label{P} \ee the Hamiltonian  $H^0$
is given by \footnote{The factor ordering in the first term is
ambiguous. I have chosen the simplest Hermitian factor ordering},
\be H^0 = {1\over 2 V_0}P{1\over  a}P + \12 V_0a \label{miniH} \ee

\bigskip

Finally, using $P = -i \partial_a$, the first approximation to the
\wdw \ equation becomes, \be -\partial_a {1\over a}\partial_a \Psi
- V_0^2a \Psi=0. \label{miniwdw} \ee The equation  has the two
solutions, \be \Psi = \exp({\pm i V_0 a^2}), \label{minipsi} \ee
 corresponding to expanding and contraction universes; to see which is which we use
(\ref{P}). The expanding solution, labeled $\Psi_0$ is \be \Psi_0
= \exp({ -i V_0 a^2}). \label{psiexpanding} \ee From now on we
will only consider this branch.

\bigskip

There is something funny about (\ref{psiexpanding}). Multiplying
$V_0$ by $a^2$ seems like an odd operation. $V_0 a^3$ is the
proper volume, but what is $V_0 a^2$? The answer in flat space is
that it is junk, but in hyperbolic space its just the proper area
of the boundary at $R_0$. One sees from the metric (\ref{FRW metric
in terms of T}) that the  coordinate volume $V_0$, and the
coordinate area $A_0$, of the boundary at $R_0$, are
(asymptotically) equal to one another, to within a factor of $2$.
\be A_0 = 2 V_0. \label{AV} \ee Thus the expression in the
exponent in (\ref{psiexpanding}) is $-\12 i A$,
 where $A$ is the
proper area  of the boundary at $R_0$. \be \Psi_0 = \exp({ -2i
A}). \label{areaphase} \ee This is a suggestive indication of a
\wdw \ boundary-holography of open FRW.

\bigskip

To go beyond the mini-superspace approximation one writes the wave
function as a product of $\Psi_0$, and a second factor $\psi(a, h,
\Phi)$ that depends on the fluctuations. \be \Psi(a, h, \Phi)=
\Psi_0 \ \ \psi(a, h, \Phi) = \exp({ -i V_0 a^2}) \ \ \psi(a, h,
\Phi). \label{fullwave} \ee By integrating the \wdw \ equation
over space, and substituting (\ref{fullwave}), an equation for
$\psi$ can be obtained. \be i\partial_a \psi + {1\over
aV_0}\partial_a {1\over a}  \partial_a \psi = H_m \psi
\label{smallwdw} \ee In this equation $H_m$ has the form of a
conventional Hamiltonian (quadratic in the momenta) for both
matter and metric fluctuations.

\bigskip

In the limit of large scale factor the term ${1\over
aV_0}\partial_a  {1\over a}
\partial_a \psi$ becomes negligible and (\ref{smallwdw}) takes the form of a Schrodinger
equation. \be i\partial_a \psi  = H_m \psi \label{schrod} \ee
Evidently the role of $a$ is not as a conventional observable, but
a parameter representing the unfolding of cosmic time. One does
not calculate its probability, but instead constrains it--perhaps
with a delta function or a Lagrange multiplier. As Banks has
emphasized \cite{banks}, in this limit, and maybe \textbf{\emph{
only }}in this limit, the wave function $\psi$ has a conventional
interpretation as a probability amplitude.
\bigskip

\subsection{Holographic WDW}

All of this brings us to the central question of this lecture: what
form does the correct holographic theory take in  asymptotically
warm cosmological backgrounds? The answer suggested in FSSY was  a
holographic version of the \wdw \ theory, living on the space-like
boundary \sig.

\bigskip

As we have described it, the \wdw \ theory is a throwback to an
older view of quantum gravity based on the existence of bulk,
space-filling degrees of freedom. It has become clear that this is
a tremendous overestimate of the capacity of space to contain
quantum information.The correct (holographic) counting of degrees
of freedom is in terms of the area of the boundary of space
\cite{hologram}. In the present case of open FRW, the special role
of the boundary is played by the surface \sig \ at $R = \infty$.

\bigskip

Just as in the ADS/CFT correspondence \cite{susskind and witten},
it is useful to define a regulated boundary, $\Sigma_0$, at
$R=R_0$. In principle $R_0$ can depend on angular location on
$\Omega_2$. In fact later we will discuss invariance under gauge
transformations of the form \be R\to R + f(\Omega_2). \label{R
shifts} \ee (the notation $ f(\Omega_2).$ indicating that $f$ is a
function of location on $\Sigma_0$.)

\bigskip

The conjecture of FSSY is that the correct Holographic description
of open FRW is a \wdw \ equation,  but  one in which the degrees
of freedom are at the boundary of space, i.e., on \sig, instead of
being distributed throughout the bulk.

\bigskip

Thus we assume the existence of a set of boundary fields, that
include a two dimensional spatial metric on $\Sigma_0$. The
induced spatial geometry of the boundary can always be described
in the conformal gauge in terms of  a Liouville  field
$U(\Omega_2).$ \be ds^2
=e^{2U(\Omega_2)}e^{2R_0(\Omega_2)}d\Omega_2^2 \label{liouville}
\ee

\bigskip

$U$ may be decomposed into  a homogeneous term $U_0$, and a
fluctuation; obviously the homogeneous term can be identified with
the FRW scale factor by \be e^{U_0} = a. \label{homogeneous part
of U} \ee In section 8 we will give a more detailed definition of
the Liouville degree of freedom.

\bigskip

 In addition we postulate  a collection of boundary ``matter" fields. The boundary matter
fields, $y$, are not the limits of the usual bulk fields $\Phi$,
but are analogous to the boundary gauge fields in the ADS/CFT
correspondence. In this paper we will not speculate on the
detailed form of  these boundary matter fields.

\bigskip

\subsection{The Wave Function}
In addition to  $U$ and $ y$, we assume a  local Hamiltonian
$H(x^i)$ that depends only on the boundary degrees of freedom (the
notation $x^i$ refers to coordinates of the boundary \sig), and a
wave function $\Psi(U, y)$, \be \Psi(U,y)= e^{-\12 S + iW}.
\label{psi equals exponential} \ee
 At every point of \sig, $\Psi$
satisfies \be H(x^i) \ \Psi(U,y). =0 \label{twoD wdw} \ee

\bigskip

 In equation (\ref{psi equals exponential}), $S(U,y)$ and $W(U,y)$ are real functionals
of the boundary fields. For reasons that will become clear, we
will call $S$ the action. However, $S$ should not in any way be
confused  with the four-dimensional Einstein action.

\bigskip

 The local Hamiltonian $H(x^i)$, and the imaginary term $W$ in the exponent, play
important roles in determining the expectation values of canonical
momenta, as well as the relation between scale factor and ordinary
time. In this paper $H$ and $W$ will play secondary roles.

\bigskip

We make the following three assumptions about $S$ and $W$:

\begin{itemize}
\bigskip

\item    Both $S$ and $W$ are invariant under conformal
transformations of $\Sigma$. This follows from the symmetry of the
background geometry: open FRW .

\item The leading (non-derivative) term in the regulated form of
$W$ is $-\12A$ where $A$ is the proper area of $\Sigma_0$,
    \be
    W = -\12 \int_{\Sigma} e^{2R_0} e^{2U} +...
    \ee
    This follows from (\ref{areaphase}).

\item  $S$ and $W$ have the form of \textbf{\emph{ local }}
 two dimensional Euclidean actions on $\Sigma$. In other words they are integrals, over
$\Sigma$, of densities that involve $U$, $y$, and their
derivatives with respect to $x^i$.

\end{itemize}
\bigskip

The first of these conditions is just a restatement of the
symmetry of the \cdl \ instanton. Later we will see that this
symmetry is spontaneously broken by a number of effects, including
the extremely interesting ``Persistence of Memory" discovered in
by Garriga, Guth and Vilenkin \cite{garriga}.

\bigskip

The second condition follows from the bulk analysis described
earlier in equation (\ref{areaphase}). It allows us to make an
educated guess about the dependence of the local Hamiltonian
$H(x^i)$ on $U$. A simple form that reproduces \ref{areaphase} \
is \be H(x) = \12 e^{-2U} \pi_U^2 - 2 e^{2U} + \ ...
\label{boundary H} \ee where $\pi_U$ is the momentum conjugate to
$U$. It is easily seen that the solution to the equation $H \Psi
=0$ has the form (\ref{areaphase}).
\bigskip

The highly nontrivial assumption is the third item--the locality
of the  action. As a rule quantum field theory wave functions are
not local in this sense. That the action $S$ is local  is far from
obvious. In our opinion it is the strongest (meaning the weakest)
of our assumptions and the one most in need of confirmation. At
present our best evidence for the locality is the discrete tower
of correlators, including a transverse, traceless, dimension-two
correlation function, described  in the next section. In
principle, much more information can be obtained from bulk
multi-point functions, continued to \sig. For example, correlation
functions of $h_{ij} $ would allow us to study the operator
product expansion of the energy-momentum tensor.

\bigskip

As I said, the assumption that $S$ is local is a very strong one,
but I mean it in a rather weak sense. One of the main points of
this lecture is that there is a natural RG flow in cosmology (see
section 6). By locality I mean only that $S$ is in the basin of
attraction of a local field theory. If it is true, locality would
imply that the  measure \be \Psi^* \Psi = e^{-S} \label{psistar
psi = exp-S} \ee has the form of a  local two dimensional
Euclidean  field theory  with action $S$, and that the Census
Taker's observations could be organized not only by conformal
invariance but by conformal field theory.

 \setcounter{equation}{0}
  \section{Data}

  The conjectured locality of the action $S$ is based on data calculated by FSSY. The
background geometry studied in \cite{yeh} was the Minkowski
continuation of a thin-wall \cdl \ instanton, describing
transitions from the Ancestor vacuum to a  hatted vacuum. For a
number of reasons such a background cannot be a realistic
description of cosmology. First of all,  there is a form of
spontaneous  breaking of  the $O(3,1)$ symmetry that Garriga,
Guth, and Vilenkin call ``The Persistence of Memory."  In section
8 we will see that  these type of effects are ``dual" to effects
expected in  the theory of RG-flows.

\bigskip

More importantly, we do not live in a universe with zero \cc.
Observational cosmology has come close to ruling out vanishing
\cc, but also theoretical considerations rule it out; in the
Landscape of String Theory the only vacua with exactly vanishing
\cc \ are supersymmetric. Nevertheless, hatted geometries are
interesting in that they are the simplest versions of
asymptotically warm geometries.

\bigskip

In FSSY, correlation functions were computed in the thin-wall,
Euclidean, \cdl \ instanton, and then continued to Minkowski
signature. The more general situation, including the possibility
of slow-roll inflation after tunneling, is presently under
investigation with Ben Freivogel, Yasuhiro Sekino, and Chen Pin
Yeh. Here we will mostly confine ourselves to the thin-wall case.

\bigskip

We begin by reviewing some facts about three-dimensional
hyperbolic space and the solutions of its massless Laplace
equation. An important distinction is between normalizable modes
(NM) and non-normalizable modes (NNM); a scalar minimally coupled
field $\chi$ is sufficient to illustrate the important points.

\bigskip

The norm in hyperbolic space is defined in the obvious way: \be \<
\chi| \chi \> = \int dR d\Omega_2 \ \chi^2 \ \sinh^2 R
\label{norm} \ee In flat space, fields that tend to a constant at
infinity are on the edge on normalizability. With the help of the
delta function,  the concept of normalizability can be generalized
to continuum-normalizability, and  the constant ``zero mode" is
included in the spectrum of the wave operator, but in hyperbolic
space the normalization integral (\ref{norm})  is exponentially
divergent for constant $\chi$. The condition for normalizability
is that $\chi \to 0$ at least as fast as $e^{-R}$. The constant
mode is therefore non-normalizable.

\bigskip

Normalizable and non-normalizable modes have very different roles in the conventional
ADS/CFT correspondence. NM are dynamical excitations with finite energy and can be produced
by events internal to the \ads. By contrast NNM cannot be excited dynamically. Shifting the value of a
NNM is equivalent to changing the boundary conditions from the bulk point of view, or changing
the Lagrangian from the boundary perspective. But, as we will see, in the
cosmological framework of FSSY,  asymptotic
warmness blurs this distinction.

\subsection{Scalars}

Correlation functions of massless (minimally coupled) scalars,
$\chi$, depend on time and on the \textbf{\emph{dimensionless}}
geodesic distance between points on ${\cal{H}}_3$. In the limit in
which the points tend to the holographic boundary \sig \ at $R \to
\infty$, the geodesic distance between points $1$ and $2$ is given
by, \be l = R_1 + R_2 + \log{(1-\cos \alpha)} \label{distance on
H} \ee where $\alpha$ is the angular distance on $\Omega_2$
between $1$ and $2$. It follows on $O(3,1)$ symmetry grounds  that
the correlation function $\< \chi(1) \chi(2) \>$ has the form,
\bea \< \chi(1) \chi(2) \> &=& G(T_1, \ T_2, \ l_{1,2}) \cr &=&
G\left\{ T_1, \ T_2, \ (R_1 + R_2 + \log{(1-\cos \alpha)})
\right\}. \label{scalar correlation in terms of T, alpha} \eea

\bigskip

Before discussing the data on the \cdl \ background, let us
consider the form of correlation functions for scalar fields in
\ads. We work in units in which the radius of the \ads \ is $1$.
By symmetry, the correlation function can only depend on $l$, the
proper distance between points. The large-distance behavior of the
two-point function has the form
   \be
   \< \chi(1) \chi(2) \> \sim {e^{-(\Delta -1)l} \over \sinh l   }.
   \label{scalar correlators in ads}
   \ee
    In anti de Sitter space the dimension $\Delta$ is related to the mass of $\chi$ by
    \be
    \Delta (\Delta - 2) = m^2.
    \label{delta m relation}
    \ee

\bigskip

    We will be interested in the limit in which the two points  $1$ and $2$ approach the
boundary at $R \to \infty$. Using (\ref{distance on H}) gives
    \be
    \< \chi(1) \chi(2) \> \sim e^{-\Delta R_1} e^{-\Delta R_2}
    {(1 - \cos \alpha)}^{-\Delta}      .
   \label{asymptotic scalar correlators in ads}
   \ee

\bigskip

    It is well known that the ``infrared cutoff" $R$, in \ads, is equivalent to an
ultraviolet cutoff in the boundary Holographic description
\cite{susskind and witten}. The exponential factors, $\exp(-\Delta
R)$ in (\ref{asymptotic scalar correlators in ads}) correspond to
cutoff dependent wave function renormalization factors and are
normally stripped off when defining boundary correlators. The
remaining factor, $ {(1 - \cos \alpha)}^{-\Delta}$ is  the
conformally covariant correlation function of a boundary field of
dimension $\Delta$.

\bigskip

In FSSY  it was claimed that in the \cdl \ background, the
correlation function contains two terms,  one of which was
associated with NM and the other with NNM. A third term was found,
but ignored on the basis that it was negligible when continued to
the boundary. In fact the third term has an interesting
significance that we will come back to, but first we will review
the terms studied in FSSY.

\bigskip

In \cite{yeh} \ the correlation function was expressed as a sum of
two contour integrals on the $k$ plane--$k$ being an eigenvalue of
the Laplacian on ${\cal{H}}_3$. The integral involves a certain
reflection coefficient ${\cal{R}}(k)$ for a Schrodinger equation,
derived from the wave equation on the \cdl \ instanton. The
contour integral is \be e^{-(T_1 + T_2)} \oint_C {dk \over 2\i}
{\cal{R}}(k) e^{-ik(T_1 + T_2)} {\left(  e^{-ikR} - e^{-ikR - 2\pi
k} \right) \over  2 \sinh{R} \sinh{k \pi} } \label{contours} \ee

\bigskip

The contours of integration are shown in Figure 6. The integrand
has poles at all imaginary values of $k$, with a double pole at
$k=i$. In addition there may be other singularities in the lower
half plane. FSSY studied only the terms coming from the upper
contour  labeled $\bf{a}$ in the figure. It contains two terms
related to NM and NNM respectively.

 \begin{figure}
\begin{center}
\includegraphics[width=12cm]{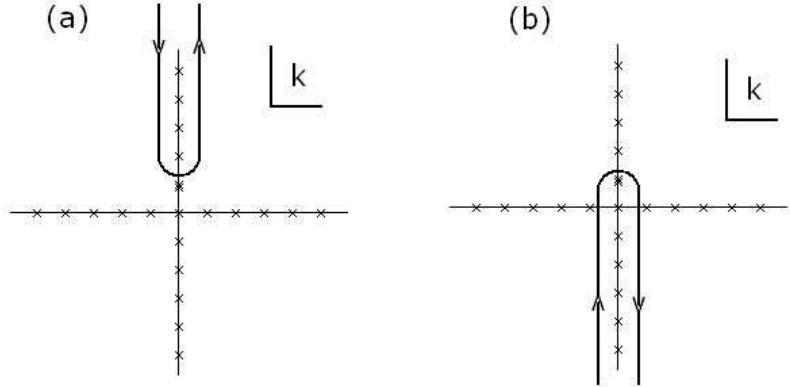}
\caption{Contours of integration for the two contributions $G_1, \
\ G_2$.} \label{6}
\end{center}
\end{figure}
 \bigskip

\bigskip

The normalizable contribution, $G_1$, is  an infinite sum, each
term having the form (\ref{asymptotic scalar correlators in ads})
with $T$-dependent coefficients. For late times, \bea
   G_1 \eq \sum_{\Delta =2}^{\infty}G_{\Delta} e^{(\Delta -2)(T_1+T_2)} {e^{-(\Delta
-1)l} \over \sinh l} \cr
  &\to& \sum_{\Delta =2}^{\infty}G_{\Delta} e^{(\Delta -2)(T_1+T_2)}  e^{-\Delta R_1}
e^{-\Delta R_2}
    {(1 - \cos \alpha)}^{-\Delta}
   \label{G1}
   \eea
   where $\Delta $ takes on integer values from $2$ to $\infty$, and $G_{\Delta}$ are a
series of constants which depend on the detailed CDL solution.

\bigskip

The connection with conformal field theory correlators is obvious;
equation (\ref{G1}) is a sum of correlation functions for fields
of definite dimension $\Delta$,  but with coefficients which
depend on the time $T$. (It should be emphasized that  the
dimensions $\Delta$  in the present context are not related to
bulk four dimensional masses by (\ref{delta m relation}).) Note
that the sum in (\ref{G1}) begins at $\Delta =2$, implying that
 every term falls at least as fast as $\exp{(-2R)}$
with respect to either argument. Thus every term is normalizable.

\bigskip

Let us now extrapolate (\ref{G1}) to the surface \sig. \sig \ can
be reached in two ways--the first being  to go out along  a
constant $T$ surface to $R = \infty$. Each term in the correlator
has a definite $R$ dependence which identifies its dimension.

\bigskip

  Another way to get to \sig \ is to first pass to light-like infinity, ${\cal{I}}^+$,
and then slide down the hat, along a light-like generator, until
reaching \sig. For this purpose it is useful to define light-cone
coordinates, $T^{\pm} = T \pm R$. \be G_1 = e^{-(T^+_1 +T^+_2)}
\sum_{\Delta} G_{\Delta} e^{(\Delta -1)(T^-_1 +T^-_2)}
(1-\cos \alpha)^{-\Delta} \label{G1 in LC coords} \ee

\bigskip

We note that apart from the overall factor $ e^{-(T^+_1 +T^+_2)}$,
$G_1$ depends only on $T^-$, and therefore tends to a finite limit
on ${\cal{I}}^+$. If we strip that factor off, then the remaining
expression consists of a sum over CFT correlators, each
proportional to a fixed power of $e^{T^-}$. In the limit ($T^- \to
- \infty$) in which we pass to \sig, each term of fixed dimension
tends to zero as $e^{(\Delta -1)(T^-_1 +T^-_2)}$ with the
dimension-2 term dominating the others.

\bigskip

The second term in the scalar correlation function discussed by
FSSY consists of a single term, \be G_2 = {e^l \over \sinh{l}   }
(T_1 +T_2 +l) \to \left\{T^+_1 + T^+_2  + \log(1 - \cos
\alpha)\right \}
 \label{G2}
\ee The contribution (\ref{G2}) does not have the form of a
correlator of a  conformal field of definite dimension.  To
understand its significance, consider a canonical massless scalar
field in two dimensions. On a two sphere the correlation function
is ultraviolet divergent and has the form \be \log\left\{\kappa^2
(1-\cos\alpha)\right\} \label{massless 2D correlator} \ee where
$\kappa$ is the ultraviolet regulator momentum. If the regulator
momentum varies with location on the sphere--for example in the
case of a lattice regulator with a variable lattice
spacing--formula (\ref{massless 2D correlator}) is replaced by \be
\log\left\{ (1-\cos\alpha)\right\} + \log \kappa_1 + \log \kappa_2
\label{massless 2D correlator with variable cutoff} \ee Evidently
if we identify the UV cutoff $\kappa$ with $T^+ $,
 \be
\log \kappa = T^+
 \label{cutoff}
 \ee
 the expressions in (\ref{G2})  and (\ref{massless 2D correlator with variable cutoff})
are identical. The relation (\ref{cutoff}) is one of the central
themes of this paper, that as we will see, relates RG flow to the
observations of the Census Taker.

 \bigskip

 That the UV cutoff of the 2D boundary theory depends on $R$ is very familiar from
the UV/IR connection \cite{susskind and witten} \ in \ads. In that
case the $T$ coordinate is absent and the log of the cutoff
momentum in the \cft \ would just be $R$. The additional time
dependent contribution in (\ref{cutoff}) will become clear later
when we discuss the Liouville field.

\bigskip

The logarithmic ultraviolet divergence in the correlator is a
signal that massless 2D scalars are ill defined; the well-defined
quantities being derivatives of the field. When calculating
correlators of derivatives, the cutoff dependence disappears. Thus
for practical purposes, the only relevant term in (\ref{massless
2D correlator with variable cutoff}) is $\log{(1-\cos\alpha)}$.

\bigskip

The existence of a dimension-zero scalar field on \sig \ is a
surprise. It is obviously associated with bulk field-modes which
don't go zero for large $R$. Such modes are non-normalizable on
the hyperbolic plane, and are usually not included among the
dynamical variables in \ads.

\bigskip

In String Theory the only massless scalars in the hatted vacua
would be moduli, which are expected to be ``fixed" in the
Ancestor. For that reason
 FSSY considered the effect of adding a four-dimensional mass term, $\mu \chi^2$, in the
Ancestor vacuum. The result on the boundary scalar was to shift
its dimension from $\Delta =0$ to $\Delta = \mu$ (for small $\mu$  less than the
Ancestor Hubble constant the corresponding mode stays non-normalizable).
However the
correlation function was not similar to those in $G_1$, each term
of  which had a  dependence on $T^-$. The dimension $\mu$ term depends
only on $T^+$:
 \be
 G_2 \to e^{-\mu T_1^+} e^{-\mu T_2^+} \log{(1 - \cos \alpha)^{-\mu}}
 \label{shifted G2}
 \ee

\bigskip

The two terms, (\ref{G1 in LC coords})  and (\ref{shifted G2})
depend on different combinations of the coordinates, $T^+$ and
$T^-$. It seems odd that there is one and only one term that
depends solely on $T^+$ and all the rest depend on $T^-$. In fact
the only reason is that FSSY ignored an entire tower of higher
dimension terms, coming from the contour $\bf{b}$ that, like
(\ref{shifted G2}), depend only on $T^+$.  From now on we will
group all  terms independent of $T^-$ into the single expression
$G_2$: \be G_2 = \sum_{\Delta'} G_{\Delta'} e^{-(\Delta' )(T^+_1
+T^+_2)}
(1-\cos \alpha)^{-\Delta'} \label{G3} \ee The $\Delta'$ include
$\mu$, the positive integers and whatever other poles appear for
$ik<1$. In the case $\mu = 0$, the leading term in $G_2$ is
(\ref{G2}).

\bigskip

We will return to the two terms $G_1$ and $G_2$  in section 6.5.

\bigskip

\subsection{Metric Fluctuations}

To prove that there is  a local field theory on \sig, the most
important test is the existence of an energy-momentum tensor. In
the ADS/CFT correspondence, the boundary energy-momentum tensor is
intimately related to the bulk metric fluctuations. We assume a
similar connection between bulk and boundary fields in the present
context. In FSSY, metrical fluctuations were studied in a
particular gauge which we will call the \textbf{\emph{ Spatially
Transverse-Traceless  }} (STT) gauge. The coordinates of region I
can be divided into FRW time, $T$, and space $x^m$ where $m = 1, \
2, \ 3.$ The STT gauge for  metric fluctuations is defined by \bea
\nabla^m h_{mn} \eq 0 \cr h_m^m  \eq 0 \label{STT gauge} \eea In
the second of equations (\ref{STT gauge}), the index is raised
with the aid of the background metric (\ref{FRW metric in terms of
T}). The main benefit of the STT gauge is that  metric
fluctuations satisfy minimally coupled, massless, scalar
equations, and the correlation functions are similar to $G_1$ and
$G_2$. However the index structure is rather involved. We define
the correlator, \bea \< h^{\mu}_{\nu} h^{\sigma}_{\tau} \> \eq G
\left\{^{\mu \sigma}_{\nu \tau}   \right\} \cr \eq G_1
\left\{^{\mu \sigma}_{\nu \tau} \right\} + G_2 \left\{^{\mu
\sigma}_{\nu \tau} \right\}. \label{gravity correlaror G1+G2} \eea

\bigskip

The complicated index structure of  $ G $ was worked out in detail
in FSSY. In this paper we quote only the results of interest--in
particular those involving  elements of $G \left\{^{\mu
\sigma}_{\nu \tau}   \right\}$ in which all indecies  lie in the
two-sphere $\Omega_2.$  Thus we consider the correlation function
$G_1 \left\{^{i k}_{j l} \right\}$.

\bigskip

As in the scalar case, $G_1 $ consists of an infinite sum of
correlators, each corresponding to a field of dimension $\Delta =
2,\ 3, \ 4,...$. The asymptotic $T$ and $R$  dependence of the
terms is identical to the scalar case, and the
first term has $\Delta = 2$. This is particulary interesting
because it is the dimension of  the energy-momentum tensor of a
two-dimensional boundary \cft. Once again this term is also time-independent.

\bigskip

After isolating the dimension-two term and stripping off the
factors $\exp{(-2R)}$, the resulting correlator is called $G_1
\left\{^{i k}_{j l}   \right\}|_{\Delta =2}$. The calculations of
FSSY \ revealed that this term is two-dimensionally
 traceless, and transverse. \bea G_1 \left\{^{i k}_{i l}
\right\}|_{\Delta =2} = G_1 \left\{^{i k}_{j k} \right\}|_{\Delta
=2} &=&0 \cr
\nabla_i G_1 \left\{^{i k}_{j l}   \right\}|_{\Delta =2} &=&0.
\label{2d transverse traceless} \eea

\bigskip

Equation (\ref{2d transverse traceless}) is the clue that, when
combined with the dimension-2 behavior of $ G_1\left\{^{i k}_{i l}
\right\}|_{\Delta =2}$, hints at a local theory on \sig. It
insures that it has the precise form of a two-point function for
an energy-momentum tensor in a conformal field theory. The only
ambiguity is the numerical coefficient connecting $G_1 \left\{^{i
k}_{j l} \right\}|_{\Delta =2}$ with $\<T^i_j T^k_l \>$. We will
return to this coefficient momentarily.

\bigskip

The existence of a transverse, traceless, dimension-two operator
is a necessary condition for the boundary theory on \sig \ to be
local: at the moment it is  our main evidence. But there is
certainly more that can be learned by computing multipoint
functions. For example, from the three-point function $\<hhh\>$ it
should be possible verify the operator product expansion and the
Virasoro algebra for the energy-momentum tensor.

\bigskip

Dimensional analysis allows us to estimate the missing coefficient
connecting the metric fluctuations with $T^i_j$, and at the same
time determine the central charge $c$.  In \cite{yeh}  \ we found
$c$ to be of order the horizon entropy of the Ancestor vacuum. We
repeat the argument here:

\bigskip

Assume that the (bulk) metric fluctuation $h$ has canonical
normalization, i.e., it has bulk mass dimension $1$ and a
canonical kinetic term. Either dimensional analysis or explicit
calculation of the two point function $\<hh\>$ shows that it is
proportional to square of the Ancestor Hubble constant. \be \< h
h\> \sim H^2. \label{hh normalization} \ee Knowing that the  three
point function $\< hhh\>$ must contain  a factor of the
gravitational coupling (Planck Length) $l_p$, it can also be
estimated by dimensional analysis. \be \< h hh\> \sim l_p H^4.
\label{hhh normalization} \ee Now assume that the 2D
energy-momentum tensor is proportional to the boundary
dimension-two part of $h$, i.e., the part that varies like
$e^{-2R}$. Schematically, \be T = q h \label{T=qh} \ee with $q$
being a numerical constant. It follows that \bea \<TT\> &\sim& q^2
H^2 \cr \<TTT\> &\sim& q^3 l_p H^4. \label{TT and TTT correlators}
\eea

\bigskip

Lastly, we use the fact that the ratio of the two and three point
functions is parametrically independent of $l_p$ and $H$ because
it is controlled by the classical algebra of diffeomorphisms:
$[T,T] = T$. Putting these elements together we find, \be \<TT\>
\sim {1 \over l_p^2 H^2} \label{final TT} \ee Since we already
know that the correlation function has the correct form, including
the short distance singularity, we can assume that the right hand
side of (\ref{final TT}) also gives the central charge. It can be
written in the rather  suggestive form: \be c \sim Area  / G  \  \
\  \  (G = l_p^2) \label{central charge} \ee where $Area$ refers
to the horizon of the Ancestor vacuum.   In other words, the
central charge of the hypothetical CFT is proportional to the
\textbf{\emph{ horizon entropy of the Ancestor  }}.

\subsection{Dimension Zero Term}

The term $G_2 \left\{^{i k}_{j l}   \right\}$ begins with a term,
which  like its scalar counterpart, has a non-vanishing limit on
\sig. It is expressed in terms of a standard 2D bi-tensor
$t\left\{^{i k}_{j l}   \right\}$ which is traceless and
transverse in the two dimensional sense. If the correlation
function were  given just by $t\left\{^{i k}_{j l} \right\}$, it
would be a pure gauge artifact. One can see this by considering
the linearized expression for the 2D curvature-scalar  $C$, \be C
= \nabla_i \nabla_j h^{ij}-2\nabla^i \nabla_i \ Tr \ h.
\label{curvature} \ee The 2D curvature associated with a traceless
transverse fluctuation vanishes, and since $t\left\{^{i k}_{j l}
\right\}$ by itself is traceless-transverse with respect to both
points, it would be  pure gauge if it appeared by itself.

\bigskip

However, the actual correlation function $G_2 \left\{^{i k}_{j l}
\right\}$  is given by \be G_2 \left\{^{i k}_{j l}   \right\} =
t\left\{^{i k}_{j l}   \right\} \left\{R_1 + T_1 + R_2 + T_2 +
\log({1 - \cos{\alpha}})\right\} \label{G2(h)} \ee The linear
terms in $R+T$, being proportional to $t\left\{^{i k}_{j l}
\right\}$ are pure gauge, but the finite term \be t\left\{^{i
k}_{j l}   \right\}
 \log({1 - \cos{\alpha}})
\label{finite part of hh} \ee gives rise to a non-trivial 2D
curvature-curvature correlation function of the form \be \<CC\> =
(1-\cos{\alpha})^{-2} \label{curvature correlator}. \ee One
difference between   the metric fluctuation $h$, and the scalar
field $\chi$, is that we cannot add a mass term for $h$ in the
Ancestor vacuum to shift its dimension.

\bigskip

Finally,  as in the scalar case, there is a tower of higher
dimension terms in the tensor correlator, $G_2 \left\{^{i k}_{j l}
\right\}$ that only depend on $T^+$.

\bigskip

The existence of a zero dimensional term in $G_2 \left\{^{i k}_{j
l}   \right\}$, which remains finite in the limit $R\to \infty$
indicates that fluctuations in the boundary geometry--fluctuations
which are due to the asymptotic warmness--cannot be ignored. One
might expect that in some way these fluctuations are connected
with the field $U$ that we encountered in the Holographic version
of the \wdw \ equation. In the next section we will elaborate on
this connection.

\bigskip\bigskip

That is the data about correlation functions on the boundary
sphere \sig \ that form the basis for our conjecture that there
exists a local holographic boundary description of the open FRW
universe. There are a number of related puzzles that  this data
raises: First, how does time emerge from a Euclidean QFT? The bulk
coordinate $R$ can be identified with scale size just as in
ADS/CFT but the origin of time requires a new mechanism.

\bigskip

The second puzzle concerns the number of degrees of freedom in the
boundary theory. The fact that the central charge is the entropy
of the Ancestor suggests that there are only enough degrees of
freedom to describe the false vacuum and not the much large number
needed for the open FRW universe at late time.

\setcounter{equation}{0}
\section{Liouville Theory}

\subsection{Breaking Free of the STT Gauge}

The existence of a Liouville sector describing metrical
fluctuations on \sig \ seems dictated by both the Holographic \wdw
\ theory and from the data of the previous section. It is clear
that the Liouville field is  somehow connected with the
non-normalizable metric fluctuations whose correlations are
contained in (\ref{G2(h)}), although the connection is somewhat
obscured by the choice of gauge in \cite{yeh}.  In the  STT gauge
the fluctuations $h$ are traceless, but not transverse (in the 2D
sense). From the viewpoint of 2D geometry they are not pure gauge
as can be seen from the fact that the 2D curvature correlation
does not vanish. One might be tempted to identify the Liouville
mode with the zero-dimension piece of (\ref{G2(h)}). To do so
would of course require a coordinate transformation on $\Omega_2$
in order to bring the fluctuation $h_i^j$ to the ``conformal" form
${\tilde{h}} \delta_i^j$.

\bigskip

This identification may be useful but it is not consistent with
the \wdw \ philosophy. The Liouville field $U$ that appears in the
\wdw \ wave function  is not tied to any specific spatial gauge.
Indeed, the wave function is required to be invariant under gauge
transformations, \be x^{\mu} \to x^{\mu} + f^{\mu}(x) \label{gauge
transform on x} \ee under which the metric transforms: \be g_{\mu
\nu} \to g_{\mu \nu} +\nabla_{\mu}f_{\nu} +\nabla_{\nu}f_{\mu}.
\label{gauge transform of g munu} \ee

\bigskip

Let's consider the effect of such transformations on the boundary
limit of $h_{ij}$. The components of $f$ along the directions in
\sig \ induce 2D coordinate transformation under which $h$
transforms conventionally. Invariance under these transformations
merely mean that the action $S$ must be a function of 2D
invariants.

\bigskip

Invariance under the shifts $f^R$ and $f^T$ are more interesting.
In particular the combination $f^+ = f^R + f^T$ generates
non-trivial transformations of the boundary metric $h_{ij} $. An
easy calculation shows that, \be h_i^j \to h_i^j +
f^+(\Omega_2)\delta_i^j. \label{trans of h under f+} \ee

\bigskip

In other words, shift transformations $f^{+}$, induce Weyl
re-scalings of the boundary metric. This prompts us to modify the
definition of  the Liouville field from \be U = T + \tilde{h}
\label{not liouville} \ee to \be U= T + \tilde{h} + f^+.
\label{Liouville = T + h + f} \ee

\bigskip

One might wonder about the meaning of  an equation such as
(\ref{Liouville = T + h + f}). The left side of of the equation is
supposed to be a dynamical field on \sig,  but the right side
contains an arbitrary function $f^+$. The point is that in the
Wheeler DeWitt formalism the wave function must be invariant under
shifts, but in the original analysis of FSSY  a specific gauge was
chosen. Thus, in order to render the wave function gauge
invariant, one must allow the shift $f^+ $ to be an integration
variable, giving it the status of a  dynamical field.

\bigskip

A similar example is familiar from ordinary gauge theories. The
analog of the \wdw \ gauge-free formalism would be the unfixed
theory in which one integrates over the time component of the
vector potential.  The analog of the STT gauge would be the
Coulomb gauge. To go from one to the other we would perform the
gauge transformation \be A_0 \to A_0 + \partial_0 \phi.
\label{gauge transform} \ee Integrating over the gauge function
$\phi$ in the path integral would restore the gauge invariance
that was given up by fixing Coulomb gauge.

\bigskip

Returning to the Liouville field, since both $\tilde{h}$ and $f$
are linearized fluctuation variables,  we see that the classical
part of $U$ is still the FRW conformal time.

\bigskip

One important point: because the effect of the shift $f^+$ is
restricted to the trace of $h$, it does not influence the
traceless-transverse (dimension-two) part of the metric
fluctuation, and the original identification of the 2D
energy-momentum tensor is unaffected.

\bigskip
Finally, invariance under the shift $f^-$ is trivial in this
order, at least for the thin wall geometry. The reason is that in
the background geometry, the area does not vary along the $T^-$
direction.

\bigskip

Given that the boundary theory is local, and includes a boundary
metric, it is constrained by the  rules of two-dimensional quantum
gravity laid down long ago by Polyakov \cite{polyakov}. Let us
review those rules for the case of a conformal ``matter" field
theory coupled to a Liouville field. Two-dimensional coordinate
invariance implies that the central charge of the Liouville sector
cancels the central charge of all other  fields. We have argued in
\cite{yeh} (and in section 5) that the central charge of the
matter sector is of order
 the horizon area of the Ancestor vacuum, measured in Planck Units. It is obvious from
the 4-dimensional bulk viewpoint that the semiclassical analysis
that we have relied on, only makes sense when  the Hubble radius
is much larger than the Planck scale. Thus we we take the central
charge of matter to satisfy $c>>1$. As a consequence, the central
charge of the Liouville sector, $c_L$, must be large and negative.
Unsurprisingly,  the negative value of $c$ is the origin of the
emergence of time.

\bigskip

The formal development of Liouville theory begins by defining two
metrics on $\Omega_2$. The first is what I will call the reference
metric ${\hat{g}}_{ij}$. Apart from an appropriate degree
smoothness, and the assumption of Euclidean signature, the
reference metric is arbitrary but fixed. In particular it is not
integrated over in the path integral. Moreover, physical
observables must be independent of ${\hat{g}}_{ij}$

\bigskip

The other metric is the ``real" metric denoted by $g_{ij}$. The
purpose of the reference metric is merely to implement a degree of
gauge fixing. Thus one assumes that the real metric has the form,
\be g_{ij} = e^{2U}{\hat{g}}_{ij}. \ee The real metric--that is to
say $U$--is a dynamical variable to be integrated over.

\bigskip

 For positive $c_L$ the Liouville Lagrangian is
 \be
L_L= {Q^2 \sqrt{\hat{g}} \over 8\pi}\left\{ \hat{\nabla}  U
\hat{\nabla}  U  + \hat{R} u \right\} \label{L-lagrangian}
 \ee
 where $ \hat{R}, \ \ \hat{\nabla} $, all refer to the  sphere
$\Omega_2$, with  metric  ${\hat{g}}$. The constant $Q$ is related
to the central charge $c_L$ by
 \be
 Q^2 = {c_L  \over 6  }
 \label{Q}
 \ee
 The two dimensional cosmological constant has been set to zero for the moment, but it
will return to play a surprising role. For future reference we
note that the cosmological term, had we included it, would have
had the form,
 \be
 L_{cc} = \sqrt{\hat{g}}\lambda e^{2U}.
 \label{Lcc}
 \ee

 \bigskip

It is useful to define a field $\phi = 4QU $ in order to bring the
kinetic term to canonical form. One finds,
 \be
L_L= {\sqrt{\hat{g}} \over 8\pi}\left\{ \hat{\nabla}  \phi
\hat{\nabla}  \phi + Q\hat{R} \phi \right\} \label{canon
L-lagrangian}
 \ee
 and, had we included a cosmological term, it  would be
  \be
 L_{cc} =\sqrt{\hat{g}} \lambda \exp{ \phi \over 2Q}.
 \label{canon Lcc}
 \ee

 \bigskip

By comparison with the case of positive $c_L$, very little is
rigorously understood about Liouville theory with negative central
charge. In this paper we will make a huge leap of faith that may
well come back to haunt us: we assume that the theory can be
defined by analytic continuation from positive $c_L$. To that end
we note that the only place that the central charge enters
(\ref{canon L-lagrangian}) and (\ref{canon Lcc}) is through the
constants $Q$ and $\gamma$, both of which become imaginary when
$c_L$ becomes negative. Let us define, \be Q = i\cal{Q}.
\label{calQ} \ee

\bigskip

Equations (\ref{canon L-lagrangian}) and (\ref{canon Lcc}) become,
\bea L_L \eq {\sqrt{\hat{g}} \over 8\pi}\left\{ \hat{\nabla}  \phi
\hat{\nabla}  \phi + i{\cal{Q}}\hat{R} \phi \right\} \cr \eq
{\sqrt{\hat{g}} \over 8\pi}\left\{ \hat{\nabla}  \phi \hat{\nabla}
\phi + 2i{\cal{Q}} \phi \right\} \label{imaginary lagrangian}
 \eea
 (where we have used $\hat{R} =2$),
 and
  \be
 L_{cc} =\sqrt{\hat{g}} \lambda \exp{-i \phi \over 2\cal{Q}}.
 \label{canon Lcc}
 \ee

\bigskip

Let us come now to the role of $\lambda$. First of all   $\lambda
$ has nothing to do with the four-dimensional cosmological
constant, either in the FRW patch or the Ancestor vacuum.
Furthermore it is not a constant in the action of the boundary
theory. Its proper
  role is as a \textbf{\emph{Lagrange multiplier}} that serves to specify the time $T$,
or more exactly, the global scale factor. The procedure is
motivated by the \wdw \ procedure of identifying the scale factor
with time. In the present case of the thin-wall limit, we identify
$\exp{2U}$ with $\exp{2T}$. Thus we insert  a $\delta$ function in
the path integral, \be \delta \left(\int  {\sqrt{\hat{g}}} (e^{2U}
- e^{2T})\right) =
\int dz \exp{iz\left(\int  {\sqrt{\hat{g}}} (e^{2U} -
e^{2T})\right)} \label{delta fnct} \ee

\bigskip

The path integral (which now includes an integration over the
imaginary 2D cosmological constant $z$) involves the action \be
L_L + L_{cc}  =
 {\sqrt{\hat{g}} \over 8\pi}\left\{ \hat{\nabla} \phi \hat{\nabla}  \phi + 2i{\cal{Q}}
\phi  + 8 \pi i z \exp{-i \phi \over 2\cal{Q}} -8\pi i z
e^{2T}\right\}
 \label{total L}
\ee

\bigskip

There is  a saddle point when the potential
 \be
V  = 2i{\cal{Q}} \phi  + 8 \pi i z \exp{-i \phi \over 2\cal{Q}}
-8\pi i z e^{2T}
 \label{V}
\ee
 is stationary; this occurs at,
\bea \exp{-i \phi \over 2\cal{Q}} \eq e^{2T} \cr z \eq i
{{\cal{Q}}^2 \over 8\pi  } e^{-2T} \label{presaddle} \eea or in
terms of the original variables, \bea e^{2U} \eq e^{2T} \cr
\lambda \eq  {{\cal{Q}}^2 \over 8\pi  } e^{-2T} \label{saddle}
\eea

\bigskip

Once $\lambda$ has been determined by (\ref{saddle}), the
Liouville theory with that value of $\lambda$ determines
expectation values of the remaining variables as functions of the
time. Thus, as we mentioned earlier, the cosmological constant is
not a constant of the theory but rather a parameter that we scan
in order to vary the cosmic time.

\bigskip
It should be noted that the existence of the saddle point
(\ref{saddle}) is peculiar to the case of negative $c$.

\bigskip

 \subsection{Liouville, Renormalization, and Correlation Functions}

 \subsection{Preliminaries}

  There are two preliminary discussions that will help us understand the application of
Liouville Theory to cosmic holography. The first is about the
ADS/CFT connection between the bulk coordinate $R$, and
renormalization-group-running of the boundary field theory. There
are three important length scales in every quantum field theory.
The first is the ``low energy scale;" in the present case the low
energy scale is the radius of the sphere which we will call $L$.

\bigskip
   The second important length is the ``bare" cutoff scale--where the underlying theory is
   prescribed.
Call it $a$. The bare input is a collection of degrees of freedom,
and an action coupling them. In a lattice gauge theory the degrees
of freedom are site and link variables, and  the couplings are
nearest neighbor to insure locality\footnote{Nearest neighbor is
common but not absolutely essential. However this subtlety is  not
important for us.} In a ferromagnet they are spins situated on the
sites of a crystal lattice.

  \bigskip

The previous  two  scales have  obvious physical meaning but the
third scale is arbitrary: a sliding scale called the
renormalization or reference scale. We  denote it by $\delta$. The
reference scale is assumed to be much smaller than $L$ and much
larger than $a$, but otherwise it is arbitrary.  It helps to keep
a concrete model in  mind. Instead of  a regular lattice,
introduce a ``dust" of points with average spacing $a$. It is  not
essential that $a$ be uniform on the sphere. Thus the spacing of
dust points
 is  a function of position, $a(\Omega_2)$.  The degrees of
freedom on the dust-points, and their nearest-neighbor couplings,
will  be left implicit.

  \bigskip

Next we introduce a second dust at larger spacing, $\delta$. The
$\delta$-dust provides the reference scale. It is  well known that
for length scales greater than $\delta$, the bare theory on the
$a$-dust can be replaced  by a renormalized theory defined on the
$\delta$-dust. The renormalized theory will typically be more
complicated, containing second, third, and $n^{th}$ neighbor
couplings.

  \bigskip

Generally, the dimensionless  form of the renormalized theory will
depend on $\delta$ in just such a way that physics  at longer
scales is exactly the  same as it was in the original theory. The
dimensionless parameters will flow as the  reference scale is
changed.

  \bigskip

If there is an infrared fixed-point, and if the bare theory is in
the basin of attraction of the fixed-point, then  as $\delta$
becomes much larger than $a$, the dimensionless parameters will
run to their fixed-point values. In that case the continuum limit
($a \to 0$) will be a conformal  field theory with $SO(3,1)$
invariance.

  \bigskip

  Similar things hold  in the theory of bulk \ads, although  in that case
   the discussion of  the  bare scale is less relevant--one might as well take the
   continuum limit $a \to 0$ from the start. In the  boundary
field theory the infrared scale is provided by the   spherical
boundary of ADS. From  the  bulk viewpoint  the boundary is at
infinite proper distance, at $R=\infty$. However, the time for a
signal to reach the boundary and be reflected back to the bulk is
finite. In that respect \ads \ behaves like a finite cavity,
requiring specific boundary conditions. To be definite, the bulk
theory is infrared regulated by replacing \sig \ with a
reference-boundary $\Sigma_0$, at finite $R$.
  Specifying the boundary conditions on $\Sigma_0$ is equivalent to specifying the
field theory parameters at scale $\delta$. In parallel with  the
field theory discussion,
 the cutoff $R$, can vary with angular position: $R = R_0(\Omega_2)$.  We can now state
the UV/IR connection by the simple identification,
  \be
  \delta(\Omega_2) = e^{-R_0(\Omega_2)}
  \label{delta-R relation}
  \ee
A useful slogan is that \textbf{\emph{``Motion along the $R$
direction is the same as renormalization-group flow."  }}

    \bigskip

    Now to the second preliminary--some observations about Lioville Theory.
Again it is helpful to  have  a concrete model. Liouville  theory
is closely connected with the theory of dense, planar, ``fishnet"
 diagrams \cite{holger} such as those which appear in large
N gauge theories, and matrix models \cite{hooft}\cite{shenker
douglas}\cite{migdal}. The fishnet plays  the role  of the bare
lattice in  the previous discussion, but now it's dynamical--we
sum over all fishnet diagrams, assuming only that the spacing (on
the sphere) is everwhere much smaller than the  sphere size, $L$.
As before, we call the  angular spacing between neighboring points
on the  sphere, $a(\Omega)$.

Each fishnet defines a metric on  the sphere. Let $d\alpha$ be a
small angular interval (measured in radians). The fishnet-metric
is defined by\footnote{Strictly speaking there is no need to
introduce a discrete fishnet lattice at the scale $\delta$. It is
sufficient to just define a continuous function $\delta
(\Omega_2)$, and from it define a reference metric by
(\ref{deltafishmetric}).}

 \be ds^2 = {d\alpha^2 \over a(\Omega)^2 }
\label{deltafishmetric} \ee

As before we introduce a reference scale $\delta$. It can also be
a fishnet, but now it is fixed, its vertices nailed down, not to
be integrated over. We continue to assume that $\delta $ satisfies
the inequalities, $a(\Omega) << \delta(\Omega)<<L$, but otherwise
it is arbitrary. The $\delta$-metric is  defined by \be
ds_{\delta}^2 = {d\alpha^2 \over \delta(\Omega)^2 }
\label{deltametric} \ee

  \bigskip

We can now define the  Liouville field $U$. All it is is the ratio
of the reference and fishnet
 scales:
\be e^U \equiv {\delta / a} \label{liou} \ee Using (\ref{liou}) \
together with $\delta = e^{-R}$, and $ds = {d\alpha \over a}$, we
see that $U$ is also given by the relation, \be ds = d\alpha \
e^{(R_0+U)}. \label{liou metric} \ee

  \bigskip

In (\ref{liou metric}) both $R_0$ and $U$ are functions of
location on $\Omega_2$, but only $U$ is dynamical, i.e., to be
integrated over.

  \bigskip

\subsection{Liouville in the Hat}

With that in mind, we return to cosmic holography, and consider
the metric on the regulated spatial boundary of FRW, $\Sigma_0$.
In the absence of fluctuations it is
$$
ds^2 = e^{2R_0} e^{2T} d^2\Omega_2.
$$

  \bigskip

In general relativity it is natural to allow both $R_0 $ and $T$
to vary over the sphere, so that \be ds^2 = e^{2R_0(\Omega_2)}
e^{2T(\Omega_2)} d^2\Omega_2 \label{metric T and R} \ee

  \bigskip

The parallel between (\ref{liou metric} and (\ref{metric T and R})
is obvious. Exactly as we might have expected from the \wdw \
interpretation, the Liouville field, $U$, may be identified with
time $T$, at least when both are large. \be U \approx T \label{u
equals T} \ee

  \bigskip

To summarize, Let's list a number of correspondences:
  \bea
\delta &\leftrightarrow& \mu^{-1} \leftrightarrow e^R  \cr \lambda
&\leftrightarrow& e^{-T} \cr a &\leftrightarrow& e^{-T^+} =
e^{-(T+R)} \label{correspondences}
  \eea

One other point about Liouville Theory: the density of vertices of
a fishnet is normally varied by changing the weight assigned to
vertices. When the fishnet is a Feynman diagram the weight is a
coupling constant $g$. It is well known that the coupling constant
and Liouville \cc are alternate descriptions of the same thing.
Either can be used to vary the average vertex density--increasing
it either by increasing $g$ or decreasing $\lambda$.  The very
dense fishnets correspond to large $U$ and therefore large FRW
time, whereas very sparse diagrams dominate the  early Planckian
era.

\subsection{Proactive and Reactive Objects in Quantum Field Theory}

There are two kinds of objects in Wilsonian renormalization that
correspond quite closely to the terms $G_1$ and $G_2$ that we have
found in the  section 5. I don't know if there is a term for the
distinction, but I will call them \textbf{\emph{``proactive}} and
\textbf{\emph{``reactive}}. Proactive objects are not quantities
that we directly measure; they are objects which go into the
definition of the theory. The best example is the exact Wilsonian
action, defined at a specific reference scale. The form
of proactive quanities depends on that reference scale,
 and so does the value of their matrix elements; indeed
  their form,  varies with $\delta$  in such a way as to
keep the physics fixed at longer distances.

\bigskip

By contrast, reactive objects are observables whose value does not
depend at all on the reference scale. They do depend on the
``bare" cutoff scale $a$ through wave function renormalization
constants, which typically tend to zero as $a \to 0$. The wave
function renormalization constants are usually stripped off when
defining a quantum field but we will find it more illuminating to
keep them.

\bigskip

The distinction between these two kinds of objects is subtle, and
is perhaps best expressed in Polchinski's version of the exact
Wilsonian renormalization group \cite{polchinski}. In that scheme,
at every scale there is a renormalized description in terms of
local defining fields $\phi(x)$, but the proactive action grows
increasingly complicated as the reference scale is lowered.

\bigskip

Consider the exact effective action defined at reference scale
$\delta$. It is given by an infinite expansion of the form \be
L_{W}(\delta) = \sum_{\Delta = 2}^{\infty} g_{\Delta}
{\cal{O}}_{\Delta} \label{wilsonian} \ee Where
${\cal{O}}_{\Delta}$ are a set of operators of dimension $\Delta$,
and $ g_{\Delta}$ are dimensional coupling constants. The
renormalization flow is  expressed in terms of  the dimensionless
coupling constants, \be \tilde{g}_{\Delta} = g_{\Delta}
\delta^{(2-\Delta )}.
 \label{dimensionless couplings}
\ee The $\tilde{g}$ satisfy RG equations,
 \be
 {d\tilde{g}\over d \log{\delta}   } = - \beta(\tilde{g}),
 \label{rg equations}
 \ee
 and at a fixed point they are constant. Thus the dimensional constants $g_{\Delta}$ in the
Lagrangian will grow with $\delta$. Normalizing them at the bare
scale $a$, in the fixed-point case we get, \be g_{\Delta} = g_a  \
\left\{{\delta  \over a}\right\}^{(\Delta -2)} \label{fixedpoint}
\ee

\be L_{W}(\delta) = \sum_{\Delta = 2}^{\infty} {\cal{O}}_{\Delta}
\left\{{\delta \over a}\right\}^{(\Delta -2)}. \label{wilsonian
expansion} \ee

\bigskip

Now consider the two point function of the effective action, $\<
L_{W}(\delta) L_{W}(\delta) \>$, evaluated at  distance scale
$L >> \delta$

\be
\< L_{W}(\delta) L_{W}(\delta)  \> =
\sum_{\Delta = 2}^{\infty}
 \< {\cal{O}}_{\Delta}
{\cal{O}}_{\Delta} \> \left({\delta \over a}\right)^{2(\Delta -2)}
\label{LL}
\ee

\bigskip

Suppose the theory is defined on a sphere of radius $L$ and we
are interested in the correlator $\< L_{W}(\delta) L_{W}(\delta)
\>$ between points separated by angle $\alpha.$
The factor $\< {\cal{O}}_{\Delta}
{\cal{O}}_{\Delta} \>$ is the two-point function of a field of dimension $\Delta$, in a theory on
the sphere of size $L$,
with an ultraviolet cutoff at the reference scale $\delta$. Accordingly it has the form
\be
\< {\cal{O}}_{\Delta}
{\cal{O}}_{\Delta} \> = \left({\delta \over L}\right)^{2\Delta } (1-\cos
{\alpha})^{-\Delta}
\label{OO}
\ee
where the two factors of $\left({\delta \over L}\right)^{\Delta }$ are the ultraviolet-sensitive
wave function renormalization constants. The final result is
 \be \< L_{W}(\delta)
L_{W}(\delta)  \> =
  \sum_{\Delta = 2}^{\infty} C_{\Delta} \left({\delta \over a}\right)^{2(\Delta -2)}
\left({\delta \over L}\right)^{2\Delta } (1-\cos
{\alpha})^{-\Delta}
 \label{wilsonwilson}
 \ee
 Note the odd
dependence of (\ref{wilsonwilson}) on the arbitrary reference
scale $\delta$. That dependence is typical of proactive
quantities.

\bigskip

Now consider a reactive quantity such as a fundamental  field,  a
derivative of such a field, or a local product of fields and
derivatives.   Their matrix elements at distance scale $L$ will be
independent of the reference scale (although it will depend of the
bare cutoff $a$) and be of order, \be \langle \phi \phi \rangle
\sim \left({a \over L}\right)^{2\Delta_{\phi}} \ee where
$\Delta_{\phi}$ is the operator dimension of $\phi$.  Thus we see
two distinct behaviors  for the scaling of correlation functions:
\be \left({\delta \over a}\right)^{2(\Delta -2)} \left({\delta
\over L}\right)^{2\Delta } \ \ \ \ \rm proactive \label{proactive}
\ee
 and
 \be
\left({a \over L}\right)^{2\Delta_{\phi}}  \ \ \ \ \rm reactive
\label{reactive}
 \ee
The formulas are more complicated away from a fixed point but the
principles are the same.

\bigskip

We note that the effective action is not the only proactive
object. The energy-momentum tensor, and various currents computed
from the effective action, will also be proactive. As we will see
these two behaviors--proactive and reactive--exactly correspond to
the dependence in (\ref{G1 in LC coords})  and (\ref{G2}).

\bigskip

Now we are finally ready to complete the discussion about the
relation between  the correlators of Section 5 and
proactive/reactive operators. Begin by noting that in ADS/CFT, the
minimally coupled massless (bulk) scalar is the dilaton, and its
associated boundary field is the Lagrangian density. It may seem
puzzling that in the present case, an entire infinite tower of
operators seems to replace, what in ADS/CFT is a single operator.
In the case of the metric fluctuations a similar tower replaces
the energy-momentum tensor. The puzzle may be stated another way.
The FRW geometry consists of an infinite number of Euclidean ADS
time slices. At what time (or what 2D cosmological constant)
should we evaluate the boundary limits of the metric fluctuations,
in order to define the energy-momentum tensor? As we will see, a
parallel ambiguity exists in Liouville theory.

\bigskip

Return now, to the three scales of Liouville Theory: the infrared
scale $L$, the reference scale $\delta$, and  the fishnet scale
$a$, with $L>> \delta >> a$. It is natural to assume that the
basic theory is defined at the bare  fishnet scale $a$  by some
collection of degrees of freedom at each lattice site, and also
specific nearest-neighbor couplings--the latter insuring locality.
Now imagine a Wilsonian integration of all degrees of freedom on
scales between the fishnet scale and the reference scale,
including the fishnet structure itself. The result will be a
proactive effective action of the type we described in equation
(\ref{wilsonian}). Moreover the correlation function of $L_{eff}$
will have the form (\ref{wilsonwilson}). But now, making the
identifications (\ref{delta-R relation}) and \be {\delta \over a}
= e^U = e^T, \label{identification} \ee we see that equation
(\ref{proactive}) for  proactive scaling becomes (for each
operator in the product) \be e^{({\Delta-2})T}e^{-\Delta R}
\label{strength} \ee This is in precise agreement with the
coefficients in the expansion (\ref{G1}). Similarly the reactive
scaling (\ref{reactive}) is $ e^{-\Delta T^+}, $ in agreement with
the properties of $G_2$.

\bigskip
It is not obvious to me exactly why the bulk fields should
correspond to proactive and reactive boundary fields in the way
that they do. I might point out the solutions to the wave equation
in the bulk are generally sums of two types of modes, \bea \chi_-
&\to&   g_-(\Omega_2) \ F_-(T^-) \ e^{-2T} \cr \chi_+ &\to&
g_+(\Omega_2) \ F_+(T^+). \label{chi plus and minus} \eea
Evidently the two types of solutions couple to objects that are
reactive and proactive under the RG flow.

\bigskip

What happens to the proactive objects if we approach \sig \ by
sending $T^+ \to \infty$ and $T_-  \to - \infty$? In this limit only
the dimension-two term survives:  exactly what we would expect if
the matter action ran toward a fixed point. All of the same things
hold true for the tensor fluctuations. Before the limit $T^- \to
-\infty$, the energy-momentum tensor consists of an infinite
number of higher dimension operators but in the limit, all  tend
to zero except for the dimension two term.

\bigskip

It should be observed that the higher dimension contributions to
$G_1 \left\{^{i k}_{j l} \right\}$ are not transverse in the
two-dimensional sense. This is to be expected: before the limit is
taken, the Liouville field does not decouple from the matter
field, and the matter energy-momentum is not separately conserved.
But if the matter theory is at a fixed point, i.e., scale
invariant, the Liouville and matter do decouple and the matter
energy-momentum should be conserved. Thus, in the limit in which
the dimension-two term dominates, it should be (and is)
transverse-traceless.

\bigskip

The RG flow is usually thought of in terms of a single independent
flow-parameter. In some versions it's the logarithm of the bare
cutoff scale, and in other formulations it's the log of the
renormalization scale. In the conventional ADS/CFT framework, $R$
can play either role. One can imagine a bare cutoff at some large
$R_0$ or one can push the bare cutoff to infinity and think of $R$
as a running renormalization scale.

However, for our purposes, it is better to keep track of both
scales. One can either think of a one-dimensional (logarithmic)
axis--we can call it the ``Wilson line"--extending from the
infrared scale to the fishnet scale $a$, or a two dimensional
$R,T$ plane. In either case
 the effective action as a function of two independent
variables. Figure 7 shows a sketch of the Wilson line and the two
dimensional plane representing the two directions $R$, and $T$.
 \bigskip
 \begin{figure}
\begin{center}
\includegraphics[width=12cm]{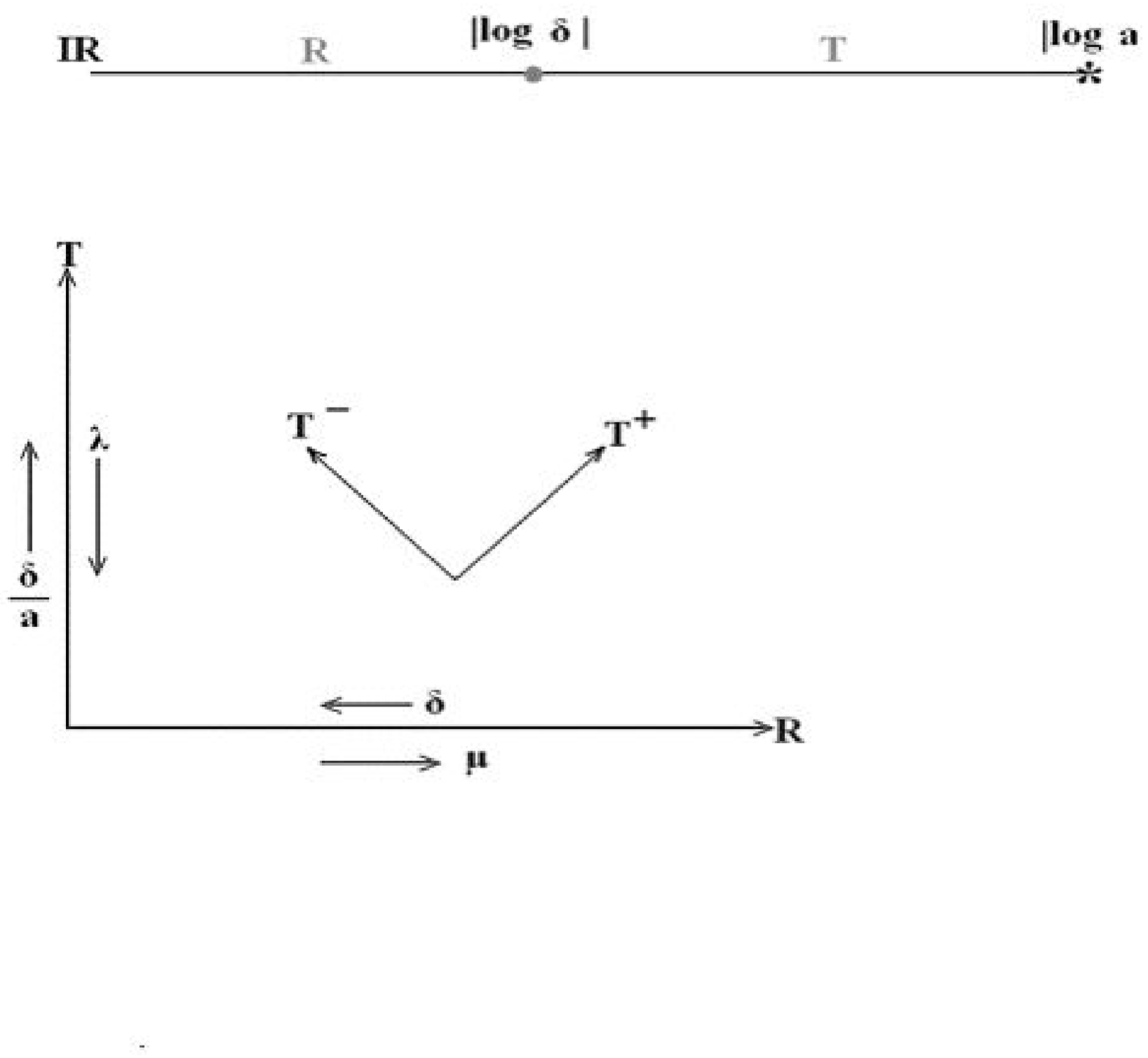}
\caption{The Wilson line of scales, and the two dimensional $R,T$
plane.} \label{7}
\end{center}
\end{figure}
 \bigskip
\bigskip

The two independent parameters can be chosen to be $a$ and
$\delta$, or equivalently $R$ and $T$. Yet another choice is to
work in momentum space. The reference energy-scale is usually
called $\mu$. \be \mu = e^R \label{mu = e^R} \ee And in the case
of negative central charge,
 the two-dimensional cosmological constant $\lambda$ can replace $T$.

\bigskip

In this light, it  is extremely interesting that the distinction
between proactive and reactive scaling, corresponds to motion
along the two light-like directions $T^-$ and $T^+$ as depicted in
figure 8.

 \begin{figure}
\begin{center}
\includegraphics[width=12cm]{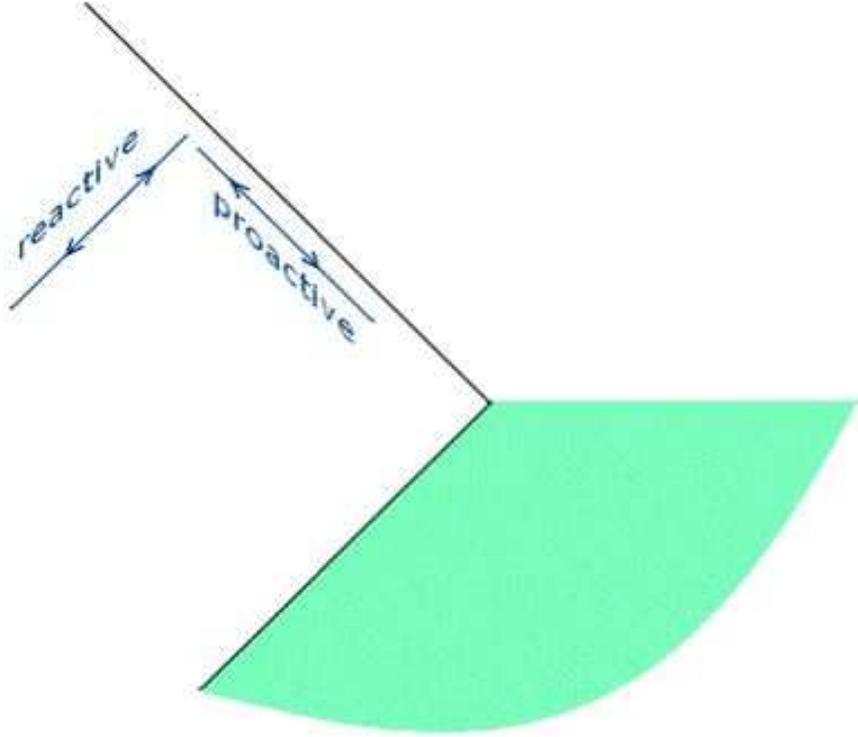}
\caption{Proactive and reactive quantities scale with the two
light-like directions $T^-$ and $T^+$}. \label{8}
\end{center}
\end{figure}
 \bigskip

It is important to understand that the duality between FRW cosmology and Liouville 2D
gravity does not only involve the continuum fixed point theory. As long as $T$ is finite
the theory has some memory of the bare theory. It's only in the limit $T \to \infty$ that the
theory flows to the fixed point and loses memory of the bare details. We will come back to this
point in section 8 when I discuss the Garriga, Guth, Vilenkin \cite{garriga}
``Persistence of Memory" phenomenon.

  \subsection{Boussovian Bounds and The c-Theorem}
When a conformal matter theory is coupled to a Liouville field,
the two sectors decouple, except for the constraint that the total
central charge vanish, \be
 c + c_L =0,
 \label{c total =0}
 \ee
 where $c$ and $c_L$
refer to matter and Liouville respectively. Moreover the matter
central charge is constant since a conformal theory is, by
definition, at a fixed-point. From now on when I speak of the
central charge I will be referring to the matter sector only.

\bigskip

There is one caveat to this rule that  $c$ is constant at a fixed
point: it applies straightforwardly as long as the reference scale
is much smaller than the infrared scale, i.e., the size of the
sphere. However, the finite (coordinate) size of the boundary
sphere provides  an infrared cutoff that is similar to the
confinement scale in a confining gauge theory. As the reference
scale becomes comparable to the total sphere, the theory runs out
of lower energy degrees of freedom and, with some definition, the
c-function will go to zero.

\bigskip

The central charge is a measure of the number of degrees of
freedom in an area-cell of size $e^{-2T^+}$. Naively, the
holographic principle would say  it is the area in such a cell in
Planck units. However, as we have seen
 in the past, this is not always the case \cite{fishcler susskind} \cite{bousso bound}.

\bigskip

Recall that motion
 up and down the $T^-$ axis, at fixed $T^+$, corresponds
to the usual RG flow of the Wilsonian action (keeping the bare
scale fixed and letting the reference scale vary). Thus, at a
fixed point,
 the area along a fixed $T^+$ line should
be constant. In the thin wall limit the area is given by \be A =
e^{2T} \ \sinh^2 R. \label{area} \ee As long as $R>>1$ the area is
indeed constant for fixed $T^+$. However as we approach $R\sim 1$
the area quickly tends to zero, consistent with the remarks above.

\bigskip

More generally, away from fixed points the Zamolodchikov c-theorem
requires $c$ to decrease with increasing reference-scale $\delta$.
This seems to suggest that the area must monotonically decrease as
$T^-$ increases, with $T^+$ held fixed, which, as we will see
raises a paradox; the area is not monotonic  beyond the thin-wall
approximation.

\bigskip

The study of how area varies along light sheets--``Boussology"--has
a long and celebrated history which I will assume you are familiar
with \cite{bousso bound}. Rather than deal with the equations that
determine
 how area varies I will draw some Bousso diagrams and tell you the conclusions. First the thin-wall
 case:
 Figure 9 shows the FRW patch of a thin-wall \cdl \ nucleation. In fact it is
 the forward light-cone of a point in flat Minkowski space. The red line is a light-sheet of constant
 $T^+$. The Bousso wedges indicate the light-like directions along which the area decreases. The
 entire geometry consists of a single region in which all wedges point toward the origin--the
 vertical side of the triangle--at $R = 0$. The area is maximum at $``enter"$: it is almost constant
 from $``enter"$ to $a$ and then $b$, but by $c$ it starts to significantly decrease. All of this
 is in accord with the c-theorem.

  \begin{figure}
\begin{center}
\includegraphics[width=12cm]{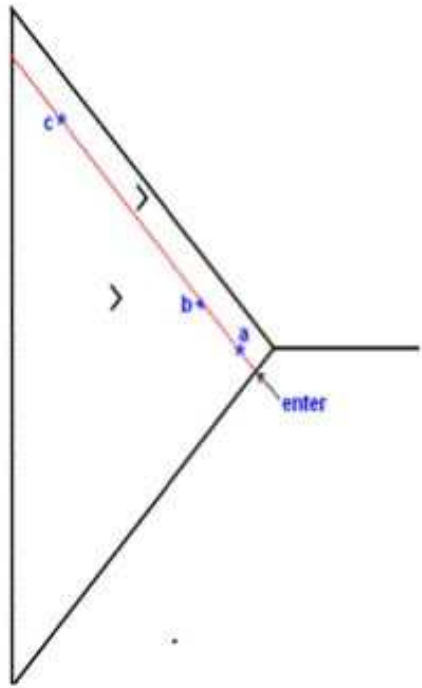}
\caption{Bousso diagram for the FRW geometry resulting from a
thin-wall CDL bubble. The entire FRW geometry consists of a single
region in which the contracting light-sheets all point in the same
directions. The light-like red line is a surface of constant
$T^+$}. \label{9}
\end{center}
\end{figure}
 \bigskip

\bigskip
 \begin{figure}
\begin{center}
\includegraphics[width=12cm]{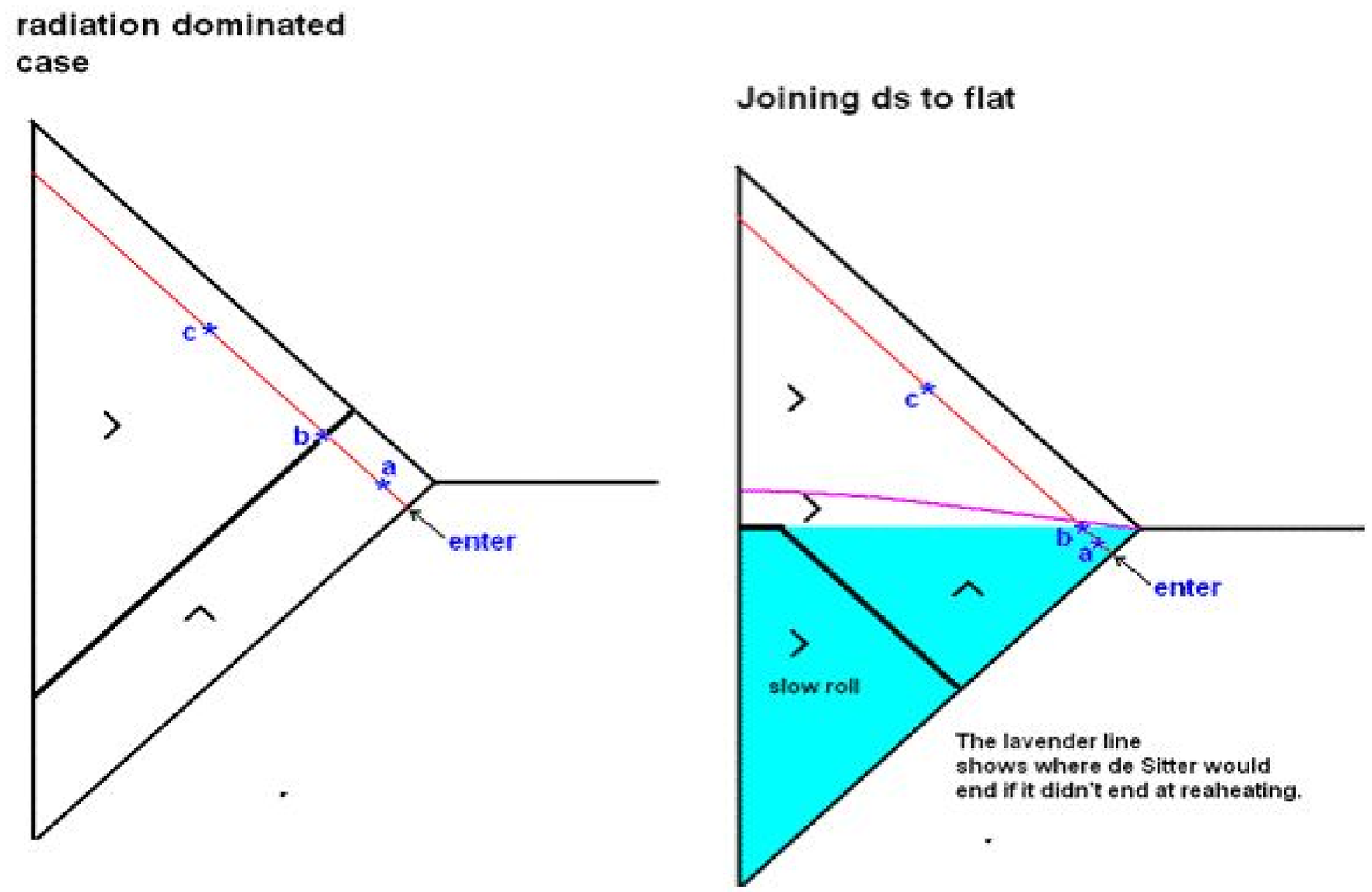}
\caption{Bousso diagrams for radiation dominated and slow-roll
universes. The slow-roll case has been simplified by attaching a
de Sitter region to a patch of flat space-time. The blue area
represents the inflating region. The light-like red line is a
surface of constant $T^+$.} \label{10}
\end{center}
\end{figure}
 \bigskip

 Next, in figure 10
some more interesting cosmological examples are shown. On the left
 is pure radiation dominated FRW. Unlike the thin-wall case it consists of two regions separated
 by the bold black line. In this case the area \textbf{\emph{increases}} from $``enter"$ to $a$,
 reaching a maximum at $b$ and then decreasing very imperceptibly to $c$. At $c$ it starts a
 quick decent to zero. This looks dangerous.

 \bigskip

 On the right side of  figure 10 the blue region is a patch with a finite vacuum energy.
 It is intended to model an era of slow-roll inflation. For simplicity, at reheating
 (horizontal edge of the blue region) I have attached it to flat space-time. More realistically
 one might attach it to the radiation dominated case but the result would be the same.

 \bigskip

 In this slow-roll case the area starts an exponential increase at $``enter"$ and again reaches
 a maximum at $b$. Beyond that the behavior is the same as for thin-wall. In fact this behavior of
 an increase of area, followed by a decrease is very generic. It would seem to violate the c-theorem.

\bigskip

 The point, however, is that the identification of $c$ with area is wrong.  The number of degrees
 of freedom of a system is a direct measure of the maximum entropy of that system. The right
 measure of $c$, for example at the point $a$, is not the area of $a$, but rather the
 maximum entropy that can pass through the light-sheet from $a$ to the origin at $R=0$. According to
 Bousso, that is given by the maximum entropy on the interval between $a$ and $b$, plus the maximum
 entropy between $b$ and the origin. Calling the areas of the various points
 $A(a), \ \ A(b), \ \ A(c)$  the maximum entropy of the light-sheet bounded by $a$ is given by,
 \be
S_{max}(a) = (A(b) - A(a) ) + (A(b)). \label{Smax}
 \ee
It is maximized at $``enter"$ and monotonically decreases to to
zero at $R=0$. Thus, identifying $c \ = \ S_{max}$ restores the
monotonicity of the central charge.

\bigskip

It's interesting to compare the naive behavior, $c \ = \ A$ with
the correct formula
 $c \ = \ S_{max}$ in the  slow-roll inflationary case. The naive formula would have $c$
 exponentially increasing from $``enter"$ to $b$. The correct formula has it exponentially
 decreasing toward its (inflated) value at $b$.

\setcounter{equation}{0}
  \section{ Scaling and the Census Taker}
  \subsection{Moments}

 Now we come to the heart of the matter: the intimate connection between the scaling
behavior of two-dimensional quantum field theory and the
observations of a Census Taker as he moves toward the Census
Bureau.
  In order to better understand the connection between the cutoff-scale  and $T^+$, let's
return to the similar connection between cutoff and the coordinate
$R$ in \ads.  We normalize the ADS radius of curvature to be $1$;
with that normalization, the Planck area is given by $1/c$ where
$c$ is the central charge.

\bigskip

  Consider the proper distance between points $1$ and $2$ given by (\ref{distance on H}).
The relation $l=R_1 +R_2 + \log{(1-\cos{\alpha})}$ is approximate,
valid when $l$ and $R_{1,2}$ are all large. When $l \sim 1$ or
equivalently, when
  \be
  \alpha^2 \sim e^{-(R_1 + R_2)}
  \label{cutoff angle large}
  \ee
equation (\ref{distance on H}) breaks down. For angles smaller
than (\ref{cutoff angle large}) the distance in \ads \ behaves
like \be l \sim e^{R}\alpha.
 \label{l small}
\ee Thus a typical correlation function will behave as a power of
$(1-\cos{\alpha})$ down to angular distances of order (\ref{cutoff
angle large}) and then fall quickly to zero.

\bigskip

The angular cutoff in \ads \ has a simple meaning. The solid angle
corresponding to the cutoff  is of order $e^{-2R}$ while the area
of the regulated boundary is $e^{+2R}$. Thus, metrically, the
cutoff area is of order unity. This means that in Planck units,
the cutoff area is the central charge $c$ of the boundary \cft.

\bigskip

Now consider the cutoff angle implied by (\ref{massless 2D
correlator with variable cutoff}) and (\ref{cutoff}). By an
argument parallel to the one above, the cutoff angle becomes
 \be
  \alpha^2 \sim e^{-(T^+_1 + T^+_2)}
  \label{cutoff angle in T+}
  \ee
Once again this corresponds to a proper area on $\Sigma_0$  (the
regulated boundary) which is time-independent, and in Planck
units,   of order the central charge.

\bigskip

Consider the Census Taker  looking back  from some late time
$T_{CT}$.   For convenience we place the CT at $R=0$. His backward
light-cone is the surface \be T + R = T_{CT} \label{census T+}.
\ee The CT can never quite see \sig. Instead he sees the regulated
surfaces corresponding to a fixed proper cutoff  (figure 4). The
later the CT observes, the smaller the  angular structure that he
can resolve on the boundary. This is another example of the UV/IR
connection, this time in a cosmological setting.

  \bigskip

\bigskip

Let's consider a specific example of a possible observation. The
massless scalar field $\chi$ of section 5 has an asymptotic limit
on \sig \ that defines the dimension zero field  $\chi(\Omega)$.
Moments of $\chi$ can be defined by integrating it with spherical
harmonics, \be \chi_{lm} = \int \chi(\Omega) Y_{lm}(\Omega)d^2
\Omega \label{moments} \ee It is worth recalling that in \ads \
the corresponding moments would all vanish because the
normalizable modes of $\chi$ all vanish exponentially as $R \to
\infty$. The possibility of non-vanishing moments is due entirely
to the asymptotic warmness of open FRW.

\bigskip

We can easily calculate \footnote{I am indebted to Ben Freivogel
for explaining equation (\ref{mean square moment}) to me.} the
mean square value of $\chi_{lm}$ (It is independent of $m$). \be
\< \chi_{l}^2 \> = \int \log(1-\cos \alpha)P_l(\cos \alpha) \sim
{1 \over l(l+1)} \label{mean square moment} \ee It is evident that
at a fixed Census Taker time $T_{CT}$, the angular resolution is
limited  by (\ref{cutoff angle in T+}). Correspondingly, the
largest moment that the CT can resolve corresponds to \be l_{max}=
e^{T_{CT}} \label{lmax} \ee Thus we arrive at the following
picture: the Census Taker can look back toward \sig \ but at any
given time his angular resolution is limited by (\ref{cutoff angle
in T+}) and (\ref{lmax}). As time goes on more and more moments
come into view. Once they are measured they are frozen and cannot
change. In other words the moments evolve from being unknown
quantum variables, with a gaussian probability distribution, to
classical boundary conditions that explicitly break rotation
symmetry (and therefore conformal symmetry). One sees from
(\ref{mean square moment}) that the  symmetry breaking is
dominated by the
 low moments.

\bigskip
But I doubt that this phenomenon ever occurs in an undiluted form.
Realistically speaking, we don't expect massless scalars in the
non-supersymmetric Ancestor. In Section 5 we discussed the effect
of a small mass term, in the ancestor vacuum, on the correlation
functions of $\chi$.
 The result of such a mass term is a
shift of the leading  dimension\footnote{In the de Sitter/CFT
correspondence \cite{DS-CFT}, the dimension of a massive scalar
becomes complex when the mass exceeds the Hubble scale. In our
case the dimension remains real for all $\mu$. I thank Yasuhiro
Sekino for this observation.} from $0$ to $\mu$. This has an
effect on the moments. The correlation function becomes \be
e^{-\mu T_1^+} e^{-\mu T_2^+} (1 - \cos{\alpha} )^{-\mu}.
\label{shifted dimension} \ee and the moments take the form
 \be
\< \chi_{l}^2 \> = e^{-2\mu T_{CT}} \int (1-\cos \alpha)^{\mu} \
P_l(\cos \alpha) \label{shifted square moment} \ee The functional
form of the $l$ dependence changes a bit, favoring  higher $l$,
but more importantly,
 the observable  effects  decrease like $e^{-2{\mu T_{CT}}}$. Thus as $T_{CT}$ advances,
the asymmetry on the sky decreases exponentially with conformal
time. Equivalently it decreases as a power of proper time along
the CT's world-line.

\bigskip

\subsection{ Homogeneity Breakdown}

\bigskip
Homogeneity in an infinite  FRW universe is generally taken for
granted, but before questioning homogeneity we should know exactly
what it means. Consider some three-dimensional scalar quantity
such as energy density, temperature, or the scalar field $\chi$.
Obviously the universe is not uniform on  small scales, so in
order to define homogeneity in a useful way we need to average
$\chi$ over some suitable volume. Thus at each point $X$ of space,
we integrate   $\chi$ over a sphere of radius $r$ and then divide
by the volume of the sphere. For a mathematically exact notion of
homogeneity the size of the sphere must tend to infinity. The
definition of the average of $\chi$ at the point $X$ is \be
\overline{\chi(X)} = \lim_{r \to \infty}{ \int \chi d^3x \over
V_r}
  \label{average}
  \ee

\bigskip
 \begin{figure}
\begin{center}
\includegraphics[width=12cm]{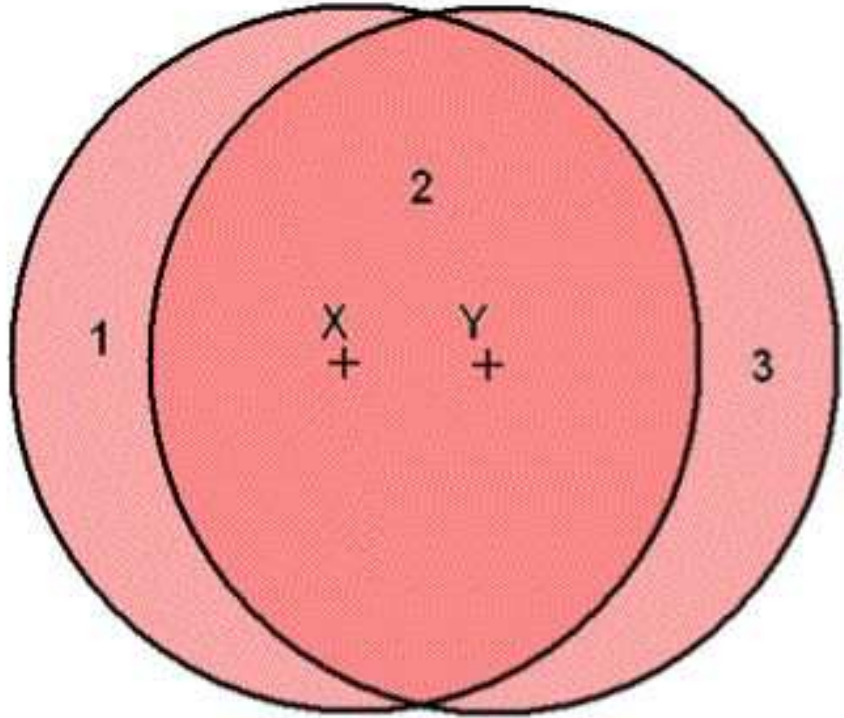}
\caption{Two large spheres centered at $X$ and $Y$.} \label{11}
\end{center}
\end{figure}
 \bigskip
Now pick a second point $Y$ and construct $\overline{\chi(Y)}$.
The difference $\overline{\chi(X)}- \overline{\chi(Y)}$ should go
to zero as $r \to \infty$ if space is homogeneous. But as  the
spheres grow  larger than  the distance between $X$ and $Y$, they
eventually almost completely overlap. In figure 11 we see that the
difference between $\overline{\chi(X)}$ and $\overline{\chi(y)}$
is due to the two thin crescent-shaped regions, 1 and 3. It seems
evident that the overwhelming bulk of the contributions to
$\overline{\chi(X)}, \ \ \ \overline{\chi(Y)}$ come from the
central region 3, which occupies almost the whole figure. The
conclusion seems to be that the averages, if they exist at all,
must be independent of position. Homogeneity while true, is a
triviality.

\bigskip

This is correct in flat space, but surprisingly it can break down
in hyperbolic space \footnote{L.S. is grateful to Larry Guth for
explaining this phenomenon, and to Alan Guth for emphasizing its
importance in cosmology.}. The reason is quite simple: despite
appearances the volume of regions 1 and 3 grow just as rapidly as
the volume of 2. The ratio of the volumes is of order \be {V_1
\over V_2} = {V_3 \over V_2} \sim  { l \over R_{curvature} } \ee
and remains finite as $r \to \infty$.

\bigskip

To be more precise we observe that \bea \overline{\chi(X)} &=& {
\int_1 \chi + \int_2 \chi  \over V_1 + V_2 } \cr
\overline{\chi(Y)} &=& { \int_3 \chi + \int_2 \chi  \over V_3 +
V_2 } \label{averages X,Y} \eea and that the difference
$\overline{\chi(X)}- \overline{\chi(Y)}$ is given by \be
\overline{\chi(X)}- \overline{\chi(Y)}= { \int_1 \chi \over V_1 +
V_2 }- { \int_1 \chi \over V_1 + V_2 } \label{difference} \ee
which, in the limit $r \to \infty$ is easily seen to be
proportional to the  dipole-moment of the boundary theory,
 \be
 \overline{\chi(X)}
-\overline{\chi(Y)} =l  \int \chi(\Omega) \cos \theta d^2 \Omega =
l \  \chi_{1,0}, \ee where $l$ is the distance between $X$ and
$Y$.

\bigskip

Since, as we have already seen for the case $\mu=0$, the mean
square fluctuation in the moments does not go to zero with
distance, it is also true that average value of
$|\overline{\chi(X)} -\overline{\chi(Y)}|^2$ will be nonzero. In
fact it grows with separation.

\bigskip

However there is no reason to believe that a dimension zero scalar
exists. Moduli, for example, are expected to be massive in the
Ancestor, and this shifts the dimension of the corresponding
boundary field. In the case in which the field $\chi$ has
dimension $\mu$, the effect (non-zero rms average of moments )
persists in a somewhat diluted form. If a renormalized field is
defined by  stripping off the wave function normalization
constants, $\exp{(-\mu T^+)}$, the squared moments still have
finite expectation values and break the symmetry. However, from an
observational point of view there does not seem to be any reason
to remove these factors. Thus it seems that as the Census Taker
time tends to infinity, the observable asymmetry will decrease
like $\exp{(-2\mu T_{CT}})$.

\bigskip

\setcounter{equation}{0}
  \section{Bubble Collisions and Other Matters}

The Census Taker idea originated with attempts to provide a
measure on the Landscape.  By looking back toward \sig, the Census
Taker can see into bubbles of other vacua--bubbles that in the
past collided with his hatted vacuum. By counting the bubbles of
each type on the sky, he can try to define a measure on the
Landscape. Whether or not this can be done, it is important to our
program to understand the representation of bubble collisions in
the language of the boundary holographic field theory.

\bigskip

Long ago, Guth and Weinberg \cite{guth weinberg} recognized that a
single isolated bubble is infinitely unlikely, and that a typical
``pocket universe" will consist of a cluster of an unbounded
number of colliding bubbles, although if the nucleation rate is
small the collisions will in some sense be rare. To see why such
bubble clusters form it is sufficient to recognize why a single
bubble is infinitely improbable.
\begin{figure}
\begin{center}
\includegraphics[width=12cm]{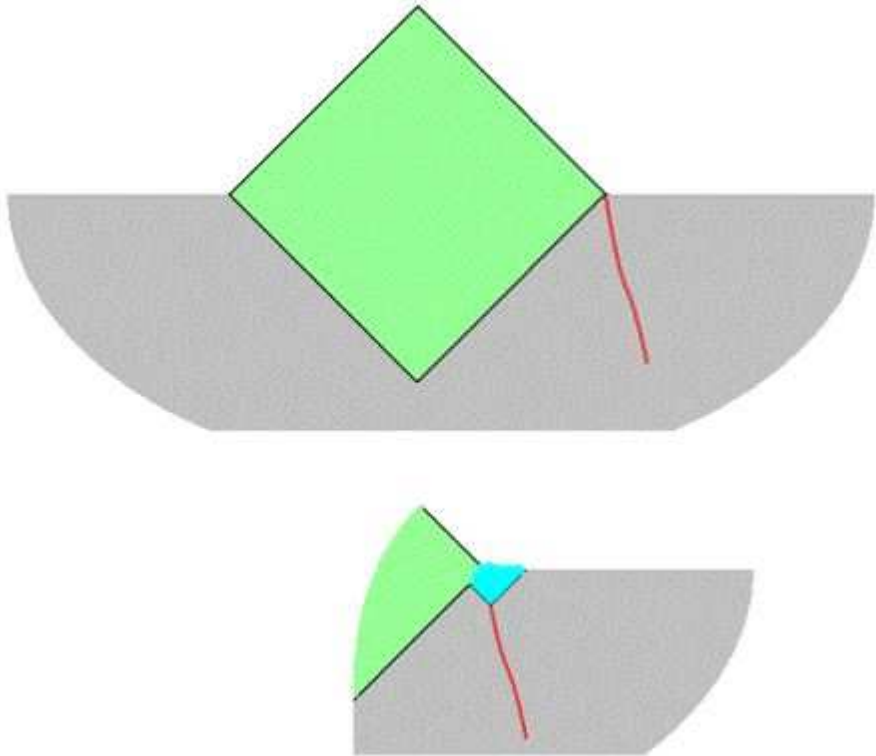}
\caption{The top figure represents a single nucleated bubble. The
red trajectory is a time-like curve of infinite length approaching
\sig. Because there is a constant nucleation rate along the curve,
it is inevitable that a second bubble will nucleate as in the
lower figure. The two bubbles will collide.} \label{12}
\end{center}
\end{figure}
In figure 12 the main point is illustrated by drawing a time-like
trajectory that approaches \sig \ from within the Ancestor vacuum.
The trajectory has infinite proper length, and assuming that there
is a uniform nucleation rate, a second bubble will eventually
swallow the trajectory and collide with the original bubble.
Repeating this  process will produce an infinite bubble cluster.

\bigskip

More recently Garriga, Guth, and Vilenkin \cite{garriga} \ have
argued that the multiple bubble collisions must spontaneously
break the $SO(3,1)$ symmetry of a single bubble, and in the
process render the (pocket) universe inhomogeneous and
anisotropic. The breaking of symmetry in \cite{garriga} was
described, not as spontaneous breaking, but as explicit breaking
due to initial conditions. However, spontaneous symmetry breaking
is nothing but the  memory of a temporary explicit symmetry
breaking, if the memory does not fade with time. For example, a
small magnetic field in the very remote past will determine the
direction of an infinite ferromagnet for all future time.
Spontaneous symmetry breaking \textbf{\emph{is}} ``The Persistence
of Memory."

\bigskip

The actual observability of bubble collisions depends on the
amount of slow-roll inflation that took place after tunneling.
Much more than 60 e-foldings would probably wipe out any signal,
but our interest in this paper is conceptual.  We will take the
viewpoint that anything within the past light-cone of the Census
Taker is in principle observable.

\bigskip

In the last section we saw that perturbative infrared effects are
capable of breaking the $SO(3,1)$ symmetry, and it is an
interesting question what the relation between these two
mechanisms is. The production of a new bubble would seem to be a
non-perturbative  effect that adds to the perturbative symmetry
breaking effects of the previous section. Whether it adds
distinctly new effects that are absent in perturbation theory is
not obvious and may depend on the specific nature of the
collision. Let us classify the possibilities.

\bigskip

\subsection{Collisions with Identical Vacua}

The simplest situation is if the true-vacuum bubble collides with
another identical bubble, the two bubbles coalescing to form a
single bubble, as in the top of figure 13.

 \begin{figure}
\begin{center}
\includegraphics[width=12cm]{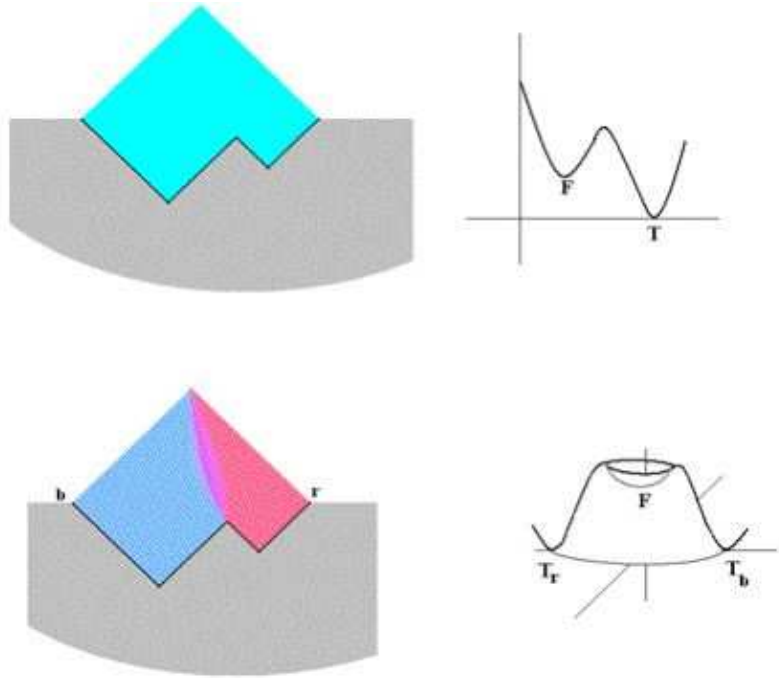}
\caption{In the top figure two identical bubbles collide. This
would be the only type of collision in a simple landscape with two
discrete minima--one of positive energy and one of zero energy. In
the lower figure a more complicated situation is depicted. In this
case the false vacuum $F$ can decay to two different true vacua,
``red"  and ``blue," each with vanishing energy. The two true
vacua  are connected by a flat direction, but CDL instantons only
lead to the red and blue points.} \label{13}
\end{center}
\end{figure}
 \bigskip

\bigskip

 The surface \sig \  is defined by starting at the tip of the hat and tracking back along
light-like trajectories until they end--in this case at a false
vacuum labeled \bf F. \rm
 The collision is parameterized by the space-like separation between nucleation points.
Particles produced at the collision of the bubbles just add to the
particles that were produced by ordinary FRW evolution. The main
effect of such a collision is to create a very distorted boundary
geometry, if the nucleation points are far apart. When they are
close the double nucleation blends in smoothly with the single
bubble. These kind of collisions seem to be no different than the
perturbative disturbances caused by the non-normalizable mode of
the metric fluctuation. Garriga, Guth, and Vilenkin, compute that
the typical observer will see multipole moments on the sky, but as
we've seen, similar multipole moments  can also occur
perturbatively.

\bigskip

In the bottom half of figure 13 we see another type of collision
in which the colliding bubbles correspond to two different true
vacua: red (r) and blue (b). But in this case red and blue are on
the same moduli-space, so that they are connected by a flat
direction\footnote{I assume that there is no symmetry along the flat
direction, and that there are only two tunneling paths from the false vacuum, one  to
red, and one to blue.}. Both vacua are included within the hat. In the bulk
space-time they bleed into each other, so that as one traverses a
space-like surface, blue gradually blends into purple and then
red.

\bigskip

On the other hand the surface \sig \ is sharply divided into blue
and red regions, as if by a one dimensional domain wall. This
seems to be  a new phenomenon that does not occur in perturbation
theory about either vacuum.

\bigskip

As an example, consider a case in which  a red  vacuum-nucleation
occurs first, and then much later a blue vacuum bubble nucleates.
In that case the blue patch on the boundary will be very small and
The Census Taker will see it occupying tiny angle on the sky. How
does the boundary field theorist interpret it? The best
description is probably as a small blue instanton in a red vacuum.
In both the bulk and boundary theory this is an exponentially
suppressed,  non-perturbative effect.

\bigskip

However, in a \cft \ the size of an instanton is a modulus that
must be integrated over. As the instanton grows the blue region
engulfs more and more of the boundary. Eventually the
configuration evolves to a blue 2D vacuum, with a tiny red
instanton. One can also think of the two configurations as the
observations of two different Census Takers at a large separation
from one  another. Which one of them is at the center, is
obviously ambiguous.

\bigskip

The same ambiguous separation into dominant vacuum, and small
instanton, can be seen another way. The nucleation sites of the
two bubbles are separated by a space-like interval. There is no
invariant meaning to say that one occurs before the other. A
element of the de Sitter symmetry group can interchange which
bubble nucleates early and which nucleates later.

\bigskip

Nevertheless, a given Census Taker will see a definite pattern on
the sky. One can always define the CT to be at the center of
things, and integrate over the relative size of the blue and red
regions. Or one can keep the size of the regions fixed--equal for
example--and integrate over the location of the CT.

\bigskip

From both the boundary field theory, and the bubble nucleation
viewpoints, the probability for any finite number of red-blue
patches is zero. Small red instantons will be sprinkled on every
blue patch and vice versa, until the boundary becomes a fractal.
The fractal dimensions are closely connected to operator
dimensions in the boundary theory. Moreover, exactly the same
pattern is expected from multiple bubble collisions.

\bigskip

But the Census taker has a finite angular resolution. He cannot
see angular features smaller than $\delta \alpha \sim
\exp{(-T_{CT})}$. Thus he will see a finite sprinkling of red and
blue dust on the sky. At $T_{CT}$ increases, the UV cutoff scale
tends to zero and the CT sees a  homogeneous``purple" fixed-point
theory.

\bigskip

The red and blue patches are reminiscent of the Ising spin system
(coupled to a Liouville field). As in that case, it makes sense to
average over small patches and define a continuous ``color field"
ranging from intense blue to intense red. It is interesting to ask
whether \sig \ would look isotropic, or whether there will be
finite multipole moments of the renormalized color color field (as
in the case of the $\chi$ field). The calculations of Garriga,
Guth, and Vilenkin suggest that multipole moments would be seen.
But unless for some reason there is a field of exactly zero
dimension, the observational signal should fade with Census Taker
time.

\bigskip

There are other types of collisions that seem to be fundamentally
different from the previous. Let us consider a model landscape
with three vacua--two false, $B$ and $W$ (Black and White); and
one true vacuum $T$. Let the the vacuum energy of   $B$ be bigger
than that of $W$, and also assume that the decays $B \to W$, $B
\to T$, and $W \to T$ are all possible. Let us also start in the
Black vacuum  and consider a transition to the True vacuum. The
result will be a hat bounded by \sig.

\bigskip

 \begin{figure}
\begin{center}
\includegraphics[width=12cm]{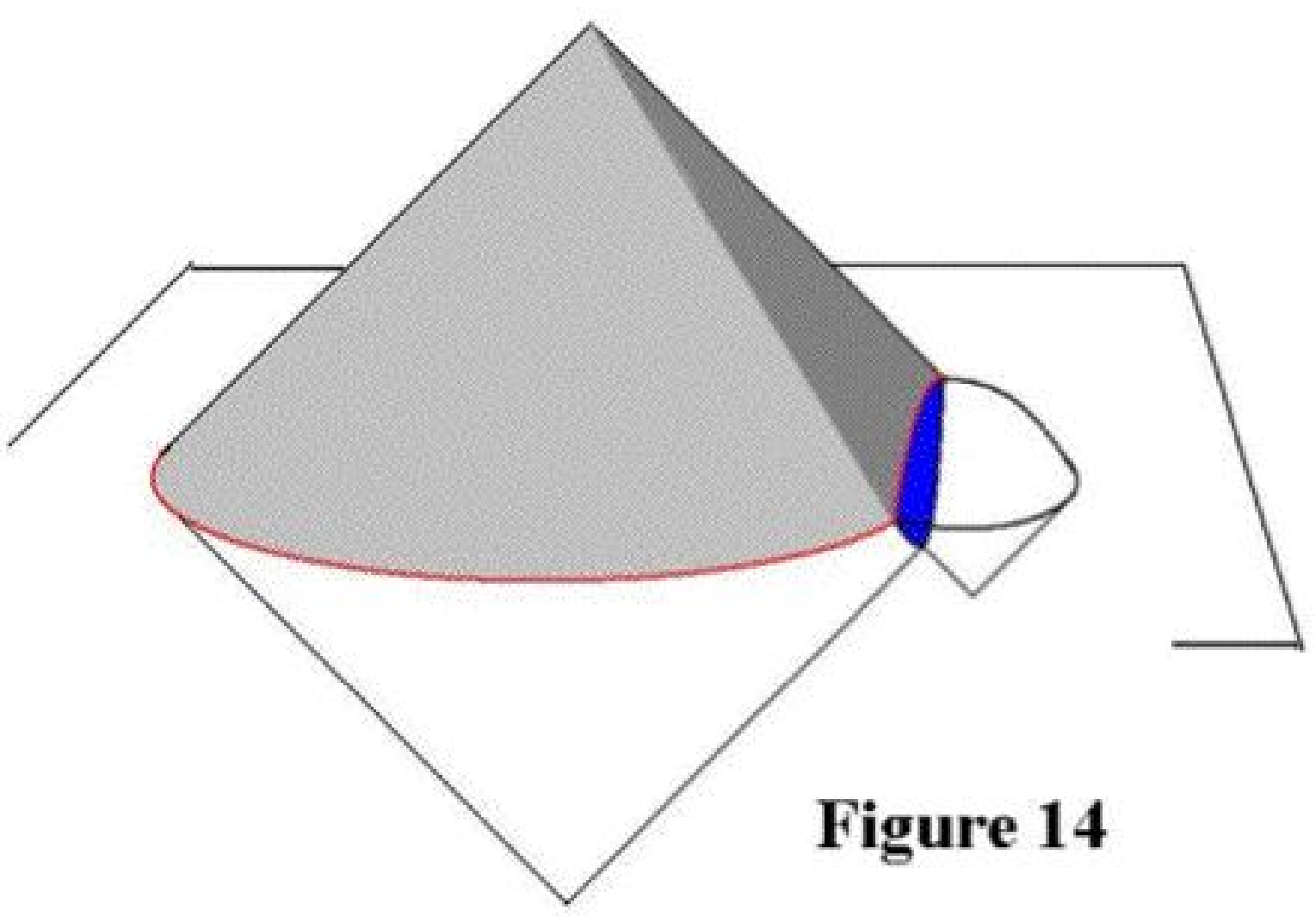}
\caption{A bubble of True vacuum forms in the Black false vacuum
and then collides with a bubble of White vacuum. The true vacuum
is bounded by a hat but the White vacuum terminates in a
space-like surface. Some generators of the hat intersect the Black
vacuum and some intersect the White. Thus \sig, shown as the red
curve, is composed of two regions. } \label{14}
\end{center}
\end{figure}
However, if a bubble of $W$ forms, it may collide with the $T$
bubble  as in figure 14. The $W$ bubble does not end in a hat but
rather, on a space-like surface. By contrast, the true vacuum
bubble does end in a hat. The surface \sig \ is defined as always,
by following the light-like generators of the hat backward until
they enter the bulk--either Black or White--as  in figure 14.

\bigskip

In this case a portion of the boundary  \sig \ butts up against
$B$, while another portion abuts $W$.
 In some
ways this situation is similar to the previous case where the
boundary was separated into red and blue regions, but there is no
analogue of the gradual bleeding of vacua in the bulk. In the
previous case the Census Taker could  smoothly pass from red to
blue.  But  in the current example, the CT would have to crash
through a domain wall in order to pass from $T$ to $W$. Typically
this happens extremely fast, long before the CT could do any
observation. In fact if we define Census Takers by the condition
that they eventually reach the Census Bureau, then they simply
never enter $W$.

\bigskip

From the field theory point of view  this example leads to a
paradox. Naively, it seems that once a $W$ patch forms on \sig, a
$B$ region cannot form inside it. A constraint of this type on
field configurations would obviously violate the rules of quantum
field theory; topologically (on a sphere) there is no difference
between a small $W$ patch  in a $B$ background, and a small $B$
patch  in a $W$ background. Thus field configurations must exist
in which a $W$ region has smaller  Black spots inside it. There is
no way consistent with locality and unitarity to forbid bits of
$B$ in regions of $W$.

\bigskip

Fortunately the same conclusion is reached from the bulk point of
view. The rules of  tunneling transitions require that if the
transition $B \to W$ is possible, so must be the transition $W \to
B$, although the probability for the latter  would be  smaller (by
a large  density of states ratio). Thus one must expect $B$  to
invade regions of $W$.

\bigskip

As the Census Taker time advances he will see smaller and smaller
spots of each type. If one assumes that there are no operators of
dimension zero, then the pattern should fade into a homogeneous
average grey,  although under the conditions I described it will
be almost White.

\bigskip

The natural interpretation is that the boundary field theory has
two phases of different free energy, the  $B$ free energy being
larger than that of  $W$. The dominant configuration would be the
ones of lower free energy with occasional fluctuations to higher
free energy.

\bigskip

\subsection{The Persistence of Memory }

The ``Persistence of Memory" reported in \cite{garriga} \ had
nothing to do with whether or not the Census Taker's sees a fading
signal: Garriga, Guth, and Vilenkin were not speaking about Census
Taker time at all. They were referring to the fact that no matter
how long after the start of eternal inflation a bubble nucleates,
it  will remember the symmetry breaking imposed by the initial
conditions; not whether the signal fades with $T_{CT}$. Returning
to figure 12, one might ask why no bubble formed along the red
trajectory in the infinitely remote past. The authors of
\cite{garriga} \ argue that eternal inflation does not make sense
without an initial condition specifying a past surface on which no
bubbles had yet formed. That surface invariably breaks the O(3,1)
symmetry and distinguishes a ``preferred Census Taker" who is at
rest in the frame of the initial surface.  He alone sees an
isotropic sky whereas all the other Census Takers see non-zero
anisotropy. What's more the effect persists no matter how late the
nucleation takes place.

\bigskip

As before, when the Census Taker's time advances, the asymmetry
should become diluted if there are no dimension zero operators,
but the existence of a preferred Census Taker at finite time makes
this  symmetry breaking seem different than what
 we have discussed up to now.

\bigskip

Let us consider  how this phenomenon  fits together with the
RG flow discussed earlier\footnote{These observations are based on
work with Steve Shenker.}. Begin by considering the behavior for
finite $\delta$ in the limit of small $a$. It is reasonable to
suppose that in integrating out the many scales between $a$ and
$\delta$, the theory would run to a fixed point. Now recall that
this is the limit of very large $T$. If in fact the theory has run
to a fixed point it will be conformally invariant. Thus we expect
that the symmetry $O(3,1)$ will be unbroken at very late time.

\bigskip

On the other hand consider the situation of $\delta/ a$ near $1$.
The reference and bare scales are very close and very few degrees
of freedom have been integrated out. There is no reason why the
effective action should be near a fixed point. The implication is
that at very early time (recall, $\delta/ a = e^T$)  the physics
on a fixed time slice will not be  conformally invariant. Near
 the beginning of an RG flow the effective action is strongly
dependent on the bare theory. The implication of a breakdown of
conformal symmetry is that there is  no symmetry between Census
Takers at different locations in space. In such situations the
center of the (deformed) \ads \ is, indeed, special.

\bigskip

Shenker and I suggest that the GGV boundary condition at the onset
of eternal inflation is the same thing as the initial condition on
the RG flow. In other words, varying the GGV boundary condition is
no different from varying the bare fishnet theory.
\bigskip

Is it possible to tune the bare action so that the theory starts
out at the fixed point? If this were so, it would be an initial
condition that allowed exact conformal invariance for all time. Of
course it would involve an infinite amount of fine tuning and is
probably not reasonable. But there may be reasons to doubt that it
is possible altogether, even though in a conventional lattice
theory it is possible.

\bigskip

The difficulty is that the bare and renormalized theories are
fundamentally different. The bare theory is defined on a variable
fishnet whose connectivity is part of the dynamical degrees of
freedom. The renormalized theory is defined on the fixed reference
lattice. The average properties of the underlying dynamical
fishnet are replaced by conventional fields on the reference
lattice. Under these circumstances it is hard to imagine what it
would mean to tune the bare theory to an exact fixed point.

\bigskip

The example of the previous subsection involving two false vacua,
$B$ and $W$,  raises some interesting questions. First imagine
starting with GGV boundary conditions such that, on some past
space-like surface, the vacuum is pure Black and that a bubble of
true vacuum nucleates in that environment. Naively the boundary is
mostly black. That means that \textbf{\emph{in the boundary
theory}} the free energy of Black must be lower than that of
White.

\bigskip

But we argued earlier that white instantons will eventually
fill \sig \ with an almost white, very light grey color, exactly
as if the initial GGV condition were White. That means that White
must have the lower potential energy. What then is the meaning of
the early dominance of $B$ from the 2D field theory viewpoint?

\bigskip

The point is that it is possible for  two rather different bare
actions to be in the same broad basin of attraction and flow to
the same fixed point. The case of Black GGV conditions corresponds
to a bare starting point (in the space of couplings) where the
potential of $B$ is lower than $W$. During the course of the flow
to the fixed point the potential changes so that at the
fixed-point $W$ has the lower energy.

\bigskip

On the other hand, White  GGV initial conditions corresponds to
starting the flow at a different bare point--perhaps closer to the
fixed point---where the potential of $W$ is lower.

\bigskip
\bigskip

This picture suggests a powerful principle. Start with the space
 of  two-dimensional actions,  which
is broad enough to contain a very large Landscape of 2D theories.
With enough fields and couplings the space could probably contain
everything. As  Wilson explained \cite{wilson}, the space divides itself
into basins of attraction. Each initial state of the universe is
described either as a GGV initial condition, or as a bare starting
point for an RG flow. The endpoints of these flows correspond to
the possible final states-the hats--that the Census Taker can end
up in.

\bigskip

\bigskip

\bigskip

We have not exhausted all the kinds of collisions that can
occur--in particular collisions with singular, negative \cc \
vacua. A particularly thorny situation results if there is a BPS
domain wall between the negative and zero CC bubbles, then as
shown by Freivogel, Horowitz, and Shenker \cite{FHS} the entire
hat may disappear in a catastrophic crunch. A possible interpretation is that
the catastrophe is due to the existence of a relevant operator which
destabilizes the fixed point. These and other issues
will be taken up in a  paper with Shenker.

\bigskip

\subsection{A Remark about Supersymmetry}

Most likely, the only 4D vacua with exactly vanishing \cc \ are
supersymmetric. Does that mean that the boundary theory on \sig \
is also supersymmetric? The answer is no: correlators on \sig \
are largely determined by the properties of the non-supersymmetric
Ancestor vacuum. For example, the gravitino will be massive in the
Ancestor, and the methods of \cite{yeh} \ would give different
dimensions, $\Delta$, for the graviton and gravitino fluctuations.

\bigskip

In fact \sig \ is contiguous with the de Sitter Ancestor and has
every reason to strongly feel the supersymmetry breaking. It is
the region close to the tip of the hat  where the physics should
be dominated by the properties of the supersymmetric terminal
vacuum. If one looks at the expansion (\ref{G1 in LC coords}), it
is clear that the tip of the hat is dominated by the
asymptotically high dimensional terms. Thus we expect
supersymmetry to manifests itself asymptotically,  in the spectrum
and operator products of  high dimensional operators.

\bigskip

\subsection{Flattened Hats and other Tragedies}

In a broad sense this paper is about phenomenology: the Census
Taker could be  us. If we lived in an ideal thin-wall hat we would
see, spread across the sky, correlation functions of a holographic
quantum field theory. We could measure the dimensions of operators
both by the time dependence of the received signals, and their
angular dependence. Bubble collisions would appear as patches
resembling instantons.

\bigskip

Unfortunately (or perhaps fortunately) we are insulated from these
effects by two forms of inflation--the slow-roll inflation that
took place shortly after bubble nucleation--and the current
accelerated expansion of the universe. The latter means that we
don't live in a true hatted geometry. Rather we live in a
flattened hat,  at least if we ignore the final decay to a
terminal vacuum.

 \begin{figure}
\begin{center}
\includegraphics[width=12cm]{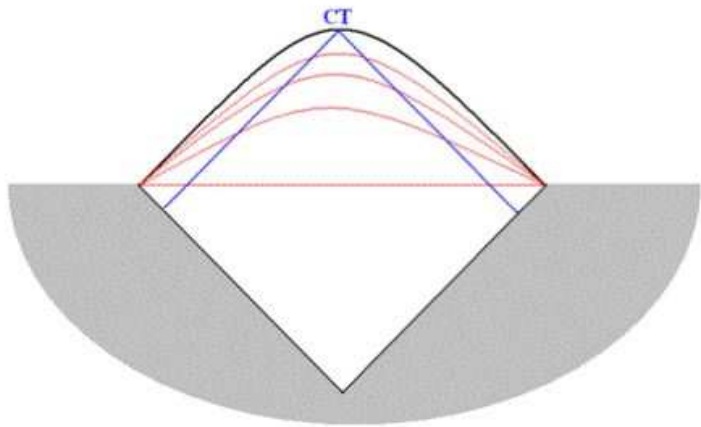}
\caption{If a CDL bubble leads to a vacuum with a small positive
cosmological constant, the hat is replaced by a rounded space-like
surface. The result is that no Census Taker can look back to
\sig.} \label{15}
\end{center}
\end{figure}
 \bigskip

The  Penrose diagram  in figure 15 shows an an Ancestor, with
large vacuum energy, decaying to a vacuum with a very small
cosmological constant. The important new feature is that the hat
is replaced by a space-like future infinity. Consider the Census
Taker's final observations as he arrives at the flattened hat. It
is obvious from  figure 18 that he cannot look back to \sig. His
past light cone is at a finite value of $T^+$. Thus for each
time-slice $T$, there is  a maximum radial variable $R = R_0(T)$
within his ken, no matter how long he waits. In other words there
is an unavoidable ultraviolet cutoff. It is completely evident
that a final de Sitter bubble must be described by a theory with
no continuum limit; in other words not only a non-local theory,
but one with no ultraviolet completion.

\bigskip

This suggests that de Sitter Space may have an intrinsic
imprecision. Indeed, as Seiberg has emphasized, the idea of a
metastable vacuum is imprecise, even in condensed matter physics
where they are common\footnote{I am grateful to Nathan Seiberg for
discussions on this point.}.

\bigskip

The more tragic fact is that all of the memory of a past bubble
may, for observational purposes, be  erased by the slow-roll
inflation that took place shortly after the \cdl \ tunneling
event--unless it lasted for the minimum permitted number of
e-foldings \cite{maria}. In principle the effects are imprinted on
the sky, but in an exponentially diluted form.

\subsection{ Note About W}

What I have described is only half the story: the half  involving
$S$, the real part of $\log{\Psi}$. The other half involves $W$,
the phase of the wave function. Knowing $S$ is enough to compute
the expectation values of the fields $y$ at a given value of time,
or, strictly speaking, at a value of the scale factor. Scanning
over scale factors can be done by varying the Liouville
cosmological constant.

\bigskip

However, quantum mechanics cannot be complete without the phase of
the wave function. In particular, the values of conjugate momenta
requires knowledge of $W$. The same is true for products of fields
at different times.

\setcounter{equation}{0}
   \section{Some Conclusions }

I've given some circumstantial evidence that there is a duality
between cosmology on the Landscape, and two-dimensional conformal
field theory, with a Liouville field. The data that supports the
theory are the computations done in \cite{yeh}, but more
importantly, a compelling physical picture accompanies the data.
The most pertinent observation is that the sky is a two-sphere.
Moreover, it is covered with interesting observable correlations;
in principle we can look back through the surface of last
scattering and observe these correlations at any past time. The
Liouville field, or alternately, the Liouville cosmological
constant (not to be confused with the four-dimensional
cosmological constant) is the dual of that time, along the
observers backward light-cone.

\bigskip

As we view the deep sky from increasingly late times, we can in
principle see greater angular detail on $\Sigma_0$, i.e., spatial
``almost-infinity". The increasing angular resolution defines a
renormalization group flow that begins with some bare action, and
ends at an infrared fixed point. The starting point
 of the flow is equivalent to the  boundary condition, whose memory, Garriga, Guth, and Vilenkin
 have argued, is persistent. The phenomena of symmetry breaking--the picking out of a special
 Census Taker at the center of things--by the GGV's initial condition,
 is equivalent to the breaking of conformal symmetry, at the start
 of a RG flow. As  in \ads,  breaking  conformal symmetry
 makes the center of ADS (at $R=0$) special.

\bigskip

Our considerations are  based on the Holographic Principle but
with a new twist. The asymptotic warmness
 of space requires a field in order to represent geometric fluctuations at infinity.
By now we are used to one or more spatial directions emerging
holographically, but this new Liouville degree of freedom creates
a new emergence--of time.

\bigskip

Examples of cosmic phenomena that can be simply interpreted as
two-dimensional field theory phenomena are the bubble collisions
of Guth and Weinberg \cite{guth weinberg}, which appear as
instantons on the two-dimensional sky; the fading of the initial
conditions with Census Taker time is connected with the spectrum
of conformal dimensions--in particular the lack of dimension zero
scalars; and ordinary slow-roll inflation corresponds to an
exponential decrease in the   Zamolodchikov c-function
\cite{c-theorem}. It will be interesting to try to interpret CMB
fluctuations in this language but this has not been studied--at
least that I know of.

\setcounter{equation}{0}
   \section{Warning to All Census Takers }

   \it
   \underline{Memo from the Director}

   \bigskip
   \bigskip
   \bigskip

   Keep in mind that when you look back toward \sig, what you see will be influenced not
only by conditions on the regulated boundary, but also by the
gravitational field between you and $\Sigma_0$. Angular
separations detected by you, must be corrected for nearby
gravitational distortions such as lensing and gravitational waves.
Please make all corrections before reporting your data.

   \rm

\bigskip

   \section{Acknowledgments}

   I am enormously  grateful to my collaborators, Ben Freivogel, Yasuhiro Sekino,
Chen-Pin Yeh, Steve Shenker, and Alex Maloney. This lecture
represents an ongoing discussion  with them. Any (positive)
reference to it
 should be accompanied by references to \cite{shenker} and
\cite{yeh}.

      \bigskip

   Discussions with Larry Guth and Alan Guth were important to my understanding of the
breaking of symmetries in Hyperbolic space.

     \bigskip

     I would also like to acknowledge helpful conversations with Raphael Bousso, another believer
     in the causal patch; Simeon Hellerman; and Matt Kleban for various insights; and finally,
     Nathan Seiberg for  explaining
     the intrinsic imprecision of metastable states.

          \bigskip

   Any conceptual or computational errors in the paper are of course my own.

\end{document}